\documentclass[12pt]{article}
\usepackage{epsf,rotate,amssymb}
\topmargin by -\headsep \textheight 8.9in \oddsidemargin 0pt
\evensidemargin \oddsidemargin \marginparwidth 0.5in \textwidth
6.5in 

\newcommand{\e}{\epsilon}
\newcommand{\bi}{\begin{itemize}}
\newcommand{\ei}{\end{itemize}}
\newcommand{\dmax}{d_{\mbox{max}}}
\newcommand{\onen}{\frac{1}{n}}

\newcommand{\AnD}{\Acal_{n\mbox{{\rm \tiny D}}}}

\newcommand{\AD}{\Acal_{\mbox{{\rm \tiny D}}}}
\newcommand{\ADSol}{\overline{\Acal^*_{\mbox{{\rm \tiny D}}}}}

\newcommand{\AnBY}{\Acal_{n\mbox{{\rm \tiny BY}}}}
\newcommand{\AoneBY}{\Acal_{1\mbox{{\rm \tiny BY}}}}
\newcommand{\ABY}{\Acal_{\mbox{{\rm \tiny BY}}}}

\newcommand{\AnS}{\Acal_{n\mbox{{\rm \tiny S}}}}
\newcommand{\AoneS}{\Acal_{1\mbox{{\rm \tiny S}}}}
\newcommand{\AS}{\Acal_{\mbox{{\rm \tiny S}}}}

\newcommand{\AnDP}{\Acal_{n\mbox{{\rm \tiny DP}}}}
\newcommand{\AoneDP}{\Acal_{1\mbox{{\rm \tiny DP}}}}
\newcommand{\ADP}{\Acal_{\mbox{{\rm \tiny DP}}}}
\newcommand{\ADPSol}{\overline{\Acal^*_{\mbox{{\rm \tiny DP}}}}}
\newcommand{\ADCSol}{\overline{\Acal^*_{\mbox{{\rm \tiny DC}}}}}

\newcommand{\AnDC}{\Acal_{n\mbox{{\rm \tiny DC}}}}
\newcommand{\AoneDC}{\Acal_{1\mbox{{\rm \tiny DC}}}}
\newcommand{\ADC}{\Acal_{\mbox{{\rm \tiny DC}}}}

\newcommand{\AnWZ}{\Acal_{n\mbox{{\rm \tiny WZ}}}}
\newcommand{\AoneWZ}{\Acal_{1\mbox{{\rm \tiny WZ}}}}
\newcommand{\AWZ}{\Acal_{\mbox{{\rm \tiny WZ}}}}

\newcommand{\AnSI}{\Acal_{n\mbox{{\rm \tiny SI}}}}
\newcommand{\AoneSI}{\Acal_{1\mbox{{\rm \tiny SI}}}}
\newcommand{\ASI}{\Acal_{\mbox{{\rm \tiny SI}}}}

\newcommand{\AcalSol}{\overline{\Acal^*}}

\newcommand{\Tenn}{{\cal T}_{\epsilon'}^{(n')}}

\newcommand{\xol}{\mbox{$\overline{x}$}}

\newcommand{\Xcalol}{\overline{\cal X}}

\newcommand{\Xcal}{{\cal X}}
\newcommand{\Ecal}{{\cal E}}
\newcommand{\Acal}{{\cal A}}
\newcommand{\Vcal}{{\cal V}}

\newcommand{\Bcal}{{\cal B}}
\newcommand{\Zcal}{{\cal Z}}

\newcommand{\Ycal}{{\cal Y}}
\newcommand{\Ucal}{{\cal U}}

\newcommand{\Yhat}{\mbox{$\hat{Y}$}}
\newcommand{\Zhat}{\mbox{$\hat{Z}$}}

\newcommand{\xhat}{\mbox{$\hat{x}$}}

\newcommand{\Fhat}{\mbox{$\hat{F}$}}

\newcommand{\Xol}{\mbox{$\overline{X}$}}

\newcommand{\Xhat}{\mbox{$\hat{X}$}}

\newcommand{\E}{\mbox{E}}
\newcommand{\be}{\begin{equation}}
\newcommand{\ee}{\end{equation}}
\newcommand{\bea}{\begin{eqnarray}}
\newcommand{\eea}{\end{eqnarray}}
\newcommand{\beann}{\begin{eqnarray*}}
\newcommand{\eeann}{\end{eqnarray*}}

\newtheorem{theorem}{Theorem}[section]

\newtheorem{lemma}[theorem]{Lemma}

\renewcommand{\theequation}{\arabic{section}.\arabic{equation}}

\newcommand{\Section}[1]{\section{#1}
\setcounter{equation}{0}
\setcounter{figure}{0}
\setcounter{table}{0}}

\title{Unified Theory of Source Coding: \\Part I -- Two Terminal Problems}

\author{Soumya Jana\\
        University of Illinois at Urbana-Champaign \\
        Email: {\tt jana@uiuc.edu}}
\date{}

\begin{document}

\maketitle

\baselineskip=1.25\normalbaselineskip
\renewcommand{\baselinestretch}{1.4}

\begin{abstract}
Since the publication of Shannon's theory of one terminal source coding, a number of interesting extensions have been derived 
by researchers such as
Slepian-Wolf, Wyner, Ahlswede-K\"{o}rner, Wyner-Ziv and Berger-Yeung. Specifically, the achievable rate or rate-distortion region has been described by a first order information-theoretic functional of the source statistics in each of the above cases. At the same time several problems have also remained unsolved. Notable two terminal examples include
the joint distortion problem, where both sources are reconstructed under a combined distortion criterion, as well as the partial side information problem, where one source is reconstructed under a distortion criterion using information about the other (side information) available at a certain rate (partially). In this paper we solve both of these open problems. Specifically, we give an infinite order description of the achievable rate-distortion region in each case. In our analysis 
we set the above problems in a general framework and 
formulate a unified methodology that solves not only the problems at hand but any two terminal problem with noncooperative encoding. The key to such unification is held by a fundamental source coding principle which we derive by extending the typicality arguments of Shannon and Wyner-Ziv. Finally, we demonstrate the expansive scope of our technique by re-deriving known coding theorems.  
We shall observe that our infinite order descriptions simplify to the expected first order in the known special cases.  
\end{abstract}

\thispagestyle{empty}
\pagestyle{plain}

\renewcommand{\baselinestretch}{1.5}

\Section{Introduction}
\label{sec:intro}
Research into source coding began with Shannon's solutions to two one-terminal problems: In the first the source is reconstructed in a lossless manner \cite{ShanLL} and in the other under a fidelity criterion \cite{Shannon}. Since then considerable effort has been directed at extending Shannon's results. 
After a hiatus of more than a decade, 
Slepian and Wolf made the next major breakthrough by finding
the achievable rate region for multiterminal lossless coding \cite{SW}. Soon after, two important discoveries took place in quick succession: 1) Wyner \cite{Wyner} as well as Ahlswede and K\"{o}rner \cite{AhlKor} derived the achievable rate region for lossless coding of one source with partial side information; 2) Wyner and Ziv gave the achievable rate-distortion region for encoding of one source with complete side information at the decoder \cite{WZ}. Quite a few interesting results have since been reported: Berger {\em et al.} derived an inner bound on the achievable rate-distortion region for encoding of one source with partial side information \cite{Upper}. Kaspi and Berger solved a source coding problem where individual encoders cooperate in a certain manner \cite{KB}. Berger and Yeung derived the achievable rate-distortion region for a two terminal problem where one source is perfectly reconstructed whereas the other is reconstructed under a fidelity criterion \cite{BY}. At this point source coding research was unmistakably geared towards solving
the famous open problem where a joint distortion criterion applies to both sources \cite{Cover}. 
Zamir and Berger took the first step in this direction by solving the joint distortion problem under a high resolution assumption \cite{Zamir}.  

In this paper, we solve the above joint distortion problem in a general setting. Specifically, we derive 
an information-theoretic expression for
the corresponding rate-distortion region as a functional of source statistics. Further, we identify the partial side information problem as a special case. Consequently, our joint distortion result solves the partial side information problem upon appropriate specialization. However, we instead present an equivalent solution that
provides specific insight into the partial side information problem. In particular, 
the inner bound derived by Berger {\em et al.} \cite{Upper} will be seen as a straightforward consequence of such specific solution.
Although we solve two hitherto open problems, the main contribution of our work lies not in providing solutions to individual problems, but in formulating a solution methodology, which (as we shall exhaustively enumerate in the second paper of this two part communication \cite{PartII}) applies to any source coding problem with noncooperating encoders and one decoder. At the heart of our method lies a fundamental principle of source coding that generalizes the typicality arguments of Shannon \cite{Shannon} and Wyner-Ziv \cite{WZ}. 
To lend credence to the versatility of our theory,
we first specialize our partial side information result 
to the case where side information is completely available at the decoder and derive Wyner-Ziv theorem as a corollary. 
We further demonstrate the expansive scope of our technique by outlining on its basis the proof of four known coding theorems given by Shannon's rate-distortion theory \cite{Shannon}, side information theory \cite{Wyner,AhlKor}, Wyner-Ziv theory (from first principle, unlike as a specialization) \cite{WZ} and Berger-Yeung theory \cite{BY}. In the second paper of this two part communication \cite{PartII}, we shall show that our methodology extends to problems involving any number of sources. 

We organize the present paper in the following manner. In Sec.~\ref{sec:prob}, we pose the joint distortion and the partial side information problems and give the respective coding theorems. We also derive Wyner-Ziv theorem by specializing the solution to the partial side information problem. In Sec.~\ref{sec:funda}, we state and prove our fundamental principle of source coding. We devote Secs.~\ref{sec:joint} and \ref{sec:partial} to the derivation of our coding theorems stated in Sec.~\ref{sec:prob}. Known coding theorems are derived based on our technique in Sec.~\ref{sec:known}. Finally, Sec.~\ref{sec:discuss} concludes the paper with a summary and the future scope of our work.

\Section{Source Coding under Joint Distortion} 
\label{sec:prob}
We begin our analysis by posing in Sec.~\ref{sec:stateJ} the two terminal joint distortion problem where the two dependent sources are separately encoded and jointly decoded under a combined distortion criterion. An information-theoretic expression of the achievable rate-distortion region is presented in Theorem \ref{th:A1}. In Sec.~\ref{sec:stateP},
we pose the partial side information problem, whose solution is presented in Theorem \ref{th:A1P}.
The inner bound of Berger {\em et al.} \cite{Upper} is then identified as a straightforward corollary of Theorem \ref{th:A1P}. We also show that, in the special case where the side information is completely available, Wyner-Ziv theorem \cite{WZ} follows from Theorem \ref{th:A1P}. First we need some notation and the notion of strong typicality.

\subsection{Notation} 

 Throughout this paper we denote random variables by uppercase letters such as $X$, $Y$, $Z$, and their alphabets by corresponding script letters ${\cal X}$, ${\cal Y}$, ${\cal Z}$. All alphabets are finite unless otherwise stated. 
By $H(X)$ and $I(X;Y)$, denote 
entropy of $X$ and mutual information between $X$ and $Y$, respectively. 
Further, denote the $k$-th element of a sequence by $x(k)$, the corresponding sequence by $\{x(k)\}$ and the collection of all elements indexed by $k_1$ through $k_2$ by $x(k_1; k_2)$. Also write $x^n=x(1;n)$, $x^n(k) = x(n(k-1)+1;nk)$ and $x^n(k_1;k_2) = x(n (k_1-1) +1; n k_2)$. In addition, denote the closure of set $\Acal$ by $\overline{\Acal}$.
Finally, define the 
$\epsilon$--strongly ($\epsilon>0$) typical set
of $X\sim p(x)$ by \cite{Cover}
\be
\label{eq:typical}
{\cal T}_\epsilon^{(n)}(X) = 
\left\{x^n\in \Xcal^n:
\left|\frac{1}{n} N(x|x^n) - p(x)\right|<\frac{\epsilon}{|\Xcal|} ~\mbox{for all}~x\in \Xcal \right\}, 
\ee
where $N(x|x^n)$ denotes the number of occurrences of $x$ in the sequence $x^n$. In this paper, we consider only strong typicality which will henceforth be mentioned as typicality.
Consequently, we have (for sufficiently large $n$) 
\be
\label{eq:TPe}
\Pr\{X^n \notin {\cal T}_\epsilon^{(n)}(X)\} \le \epsilon
\ee 
due to strong law of large numbers \cite{Cover}, 
where $\{X(k)\}$ are drawn i.i.d. $\sim p(x)$. Also if $x^n \in {\cal T}_\epsilon^{(n)}(X)$, then we call $x^n$ a typical sequence. In an analogous manner, the jointly typical set of a collection of random variables $\Xol = (X_1,X_2,...,X_M)$
is defined by (\ref{eq:typical}) with $X$, $x$ and $\Xcal$ replaced by $\Xol$, $\xol$ and $\Xcalol=\Xcal_1\times\Xcal_2\times...\times\Xcal_M$, respectively. 

\subsection{Joint Distortion}
\label{sec:stateJ}

{\bf {\em Problem Statement}:} Let  
$(X_1,X_2)\sim p(x_1,x_2)$ be two discrete random variables and draw  
i.i.d. copies $\{(X_{1}(k),X_2 (k))\}$ $\sim p(x_1,x_2)$. 
Encode 
$(X_1,X_2)$ using two encoder mappings
\beann
&&f_{1}:\Xcal_1^n \rightarrow \Zcal_1\\
&&f_{2}:\Xcal_2^n \rightarrow \Zcal_2
\eeann
for some alphabets $\Zcal_1$ and $\Zcal_2$, and
decoding using decoder mapping
$$
g: \Zcal_1\times\Zcal_2 \rightarrow \Xcal_1^n\times \Xcal_2^n
$$
under a  bounded distortion criterion
\[
d: \Xcal_1^2\times\Xcal_2^2\rightarrow [0,\dmax]
\]
defined on $\Xcal_1\times\Xcal_2$. 
We call the mapping triplet $(f_1,f_2,g)$ a joint distortion 
code of length $n$ and 
\be
\label{eq:rec}
(\Xhat_1^n,\Xhat_2^n) =g(f_1(X_1^n),f_1(X_1^n))
\ee
the estimate or reconstruction of $(X_1^n,X_2^n)$.
A rate-distortion triplet $(R_1,R_2,D)$ is said to be achievable if for any $\e>0$ there exists (for $n$ sufficiently large) a joint distortion code $(f_1,f_2,g)$  such that 
\bea
\label{eq:AR1}
\frac{1}{n} \log |\Zcal_1 | &\le& R_1 +\e \\
\label{eq:AR2}
\frac{1}{n} \log |\Zcal_2 | &\le& R_2 +\e \\
\label{eq:ALC}
\frac{1}{n} \E d_{n}((X_1^{n},X_2^{n}), (\Xhat_1^{n},\Xhat_2^n)) &\le& D +\e
\eea
where
\be
\label{eq:dn}
d_{n}((x_1^{n},x_2^{n}), (\xhat_1^{n},\xhat_2^n)) = \sum_{k=1}^n d((x_{1}(k),x_{2}(k)), (\xhat_{1}(k),\xhat_{2}(k))).
\ee
Denote the set of achievable triplets $(R_1,R_2,D)$ by $\AD$ (subscript `D' indicating a distortion criterion involved). Clearly, $\AD$ is closed.
 Our task is to express $\AD$ as an information-theoretic functional of the source distribution $p$. 

{\bf {\em Results}:} Let $\AnD^*$ be the set of $(R_1,R_2,D)$ triplets such that there exist alphabets $\Zcal_1$ and $\Zcal_2$, conditional distributions $q_1(z_1|x_1^n)$ and $q_2(z_2|x_2^n)$, and mapping $\psi:\Zcal_1\times\Zcal_2 \rightarrow \Xcal_1^n\times\Xcal_2^n$, satisfying
\bea
\label{eq:R1}
\onen
I(X_1^n;Z_1|Z_2) &\le& R_1 \\
\label{eq:R2}
\onen
I(X_2^n;Z_2|Z_1) &\le& R_2 \\
\label{eq:R12}
\onen
I(X_1^n,X_2^n;Z_1,Z_2) &\le& R_1 + R_2 \\
\label{eq:D12}
\onen \E d_{n}((X_1^n,X_2^n),\psi(Z_1,Z_2))&\le& D
\eea
where 
\be
\label{eq:pLCPC}
(X_1^n,X_2^n,Z_1,Z_2) \sim 
p_n(x_1^n,x_2^n)
q_{1}(z_1|x_1^n) q_{2}(z_2|x_2^n)
\ee 
and $p_n(x_1^n,x_2^n) =\prod_{k=1}^n p(x_{1}(k),x_{2}(k))$. Note that (\ref{eq:pLCPC}) is equivalent to stating that $Z_1 \rightarrow X_1^n \rightarrow X_2^n \rightarrow Z_2$ forms a Markov chain. Further, let 
\be
\label{eq:A*}
\AD^* = \bigcup_{n=1}^\infty \AnD^*.
\ee
Note that each $\AnD^*$ is closed; however, $\AD^*$ not necessarily is. 

\begin{theorem}
\label{th:A1}
$\AD = \ADSol$.
\end{theorem}

Next we pose the partial side information problem considered by Berger {\em et al.} \cite{Upper}.

\subsection{Partial Side Information Problem}
\label{sec:stateP}

{\bf {\em Problem Statement}:} Consider the problem where source $X_1$ is encoded with side information $X_2$ partially available at the decoder. The problem setting remains the same as that in Sec.~\ref{sec:stateJ} except that the distortion criterion $d: \Xcal_1^2\rightarrow [0,\dmax]$ is now defined over $\Xcal_1$ alone. Accordingly, a partial side information code consists of a mapping triplet 
$$(f_1:\Xcal_1^{n} \rightarrow \Zcal_1,f_2:\Xcal_2^{n} \rightarrow \Zcal_2,g:\Zcal_1\times\Zcal_2 \rightarrow \Xcal_1^{n}).$$
Denote by $\ADP$ the set of rate-distortion triplets $(R_1,R_2,D)$ achievable by such codes (additional subscript `P' indicating partial side information). Any $(R_1,R_2,D)\in \ADP$ if for any $\e>0$ there exists (for $n$ sufficiently large) partial side information code $(f_1,f_2,g)$ such that (\ref{eq:AR1}) and (\ref{eq:AR2}) hold  alongside the distortion condition
\be
\label{eq:ALC'}
\frac{1}{n} \E d_{n}(X_1^{n}, \Xhat_1^{n}) \le D +\e
\ee
(in place of (\ref{eq:ALC}) seen in case of $\AD$). 
Here $\Xhat_1^n =g(f_1(X_1^n),f_1(X_1^n))$ now. 
As earlier, note that $\ADP$ is closed.
We give an information-theoretic description of $\ADP$ in the following. 

{\bf {\em Results}:} We can specialize
Theorem \ref{th:A1} in an intuitive way to the partial side information problem where we have ($d: \Xcal_1^2\rightarrow [0,\dmax]$, $g:\Zcal_1\times\Zcal_2 \rightarrow \Xcal_1^{n}$). Specifically, define $\AnDP'^*$ in the same manner as $\AnD^*$ except that the mapping $\psi$ is now of the form 
$\Zcal_1\times\Zcal_2 \rightarrow \Xcal_1^n$ and condition (\ref{eq:D12}) is now replaced by 
\be
\label{eq:D12'}
\onen \E d_{n}(X_1^n,\psi(Z_1,Z_2))\le D.
\ee
Further, denoting $\ADP'^* = \bigcup_{n=1}^\infty \AnDP'^*$, we can show $\ADP = \ADP'^*$ (mimicking the proof of Theorem \ref{th:A1}). According to this description, any $(R_1,R_2,D)\in \ADP$ satisfies four conditions (\ref{eq:R1})--(\ref{eq:R12}) and (\ref{eq:D12'}) for some $n$. 

We can also
describe $\ADP$ in an equivalent but more insightful manner using only three conditions.
Let $\AnDP^*$ be the set of $(R_1,R_2,D)$ triplets such that there exist alphabets $\Zcal_1$ and $\Zcal_2$, conditional distributions $q_1(z_1|x_1^n)$ and $q_2(z_2|x_2^n)$, and mapping $\psi:\Zcal_1\times\Zcal_2 \rightarrow \Xcal_1^n$, satisfying
\bea
\label{eq:R1P}
\onen
I(X_1^n;Z_1|Z_2) &\le& R_1 \\
\label{eq:R2P}
\onen
I(X_2^n;Z_2) &\le& R_2 \\
\label{eq:D1P}
\onen \E d_{n}(X_1^n,\psi(Z_1,Z_2))&\le& D
\eea
where 
$(X_1^n,X_2^n,Z_1,Z_2) \sim 
p_n(x_1^n,x_2^n)
q_{1}(z_1|x_1^n) q_{2}(z_2|x_2^n)$. 
Further, let 
\be
\label{eq:AP*}
\ADP^* = \bigcup_{n=1}^\infty \AnDP^*
\ee
as earlier. Note that each $\AnDP^*$ is closed; however, $\ADP^*$ not necessarily is. 

\begin{theorem}
\label{th:A1P}
$\ADP = \ADPSol$.
\end{theorem}

In view of definition (\ref{eq:AP*}),
Theorem \ref{th:A1P} implies $\ADP \supseteq \AoneDP^*$. This inner bound was derived by Berger {\em et al.} \cite{Upper}. 
Moreover, if the side information $X_2$ is completely available at the decoder, i.e., an additional constraint $R_2 = H(X_2)$ is imposed, then it can be shown from
Theorem \ref{th:A1P} that
the set $\ADC$ (`C' in the subscript indicating complete side information) of achievable pairs $(R_1,D)$ is given by Wyner-Ziv theorem \cite{WZ}. An outline of the proof is given below. Additional steps necessary to make the proof rigorous appear in Appendix \ref{app:specialWZ}. 

{\bf {\em Wyner-Ziv Theorem as Corollary}:} Clearly, the achievable set $\ADC$ is given by
\be
\label{eq:ADCz}
\ADC = \{(R_1,D):(R_1,R_2,D)\in \ADP, R_2 = H(X_2)\}. 
\ee
Also, define $\AnDC^*$ by (\ref{eq:ADCz}) with $\ADP$ in place of $\AnDP^*$. 
Further, defining 
$\ADC^* = \bigcup_{n=1}^\infty \AnDC^*$, we have $\ADC=\ADCSol$
due to Theorem \ref{th:A1P}. 

Now let us inspect $\AnDC^*$ closely. Noting
$\onen
I(X_2^n;Z_2) \le \onen H(X_2^n) = H(X_2)$, condition (\ref{eq:R2P}) 
automatically holds for $R_2=H(X_2)$
 and, as we shall see in Appendix \ref{app:WZ1}, one may assume $Z_2=X_2^n$ with probability one without loss of generality. Hence
referring to (\ref{eq:R1P}) and (\ref{eq:D1P}), $\AnDC^*$ is now described by the set of $(R_1,D)$ pairs such that    
there exist alphabet $\Zcal_1$, conditional distribution $q_1(z_1|x_1^n)$ and mapping $g:\Zcal_1\times\Xcal_2^n \rightarrow \Xcal_1^n$, satisfying
\bea
\label{eq:R1C}
\onen
I(X_1^n;Z_1|X_2^n) &\le& R_1 \\
\label{eq:D1C}
\onen \E d_{n}(X_1^n,g(Z_1,X_2^n))&\le& D
\eea
where 
$(X_1^n,X_2^n,Z_1) \sim 
p_n(x_1^n,x_2^n)
q_{1}(z_1|x_1^n)$. Further, referring to (\ref{eq:R1C}) and (\ref{eq:D1C}), recall that
Wyner-Ziv theorem states $\ADC = \AoneDC^*$ \cite{WZ}. Consequently, since $\ADC=\ADCSol$ by Theorem \ref{th:A1P} (as noted earlier), showing $\ADCSol = \AoneDC^*$ amounts to proving Wyner-Ziv theorem.
Moreover, noting $\ADCSol \supseteq \AoneDC^*$ from definition, we only require $\ADCSol \subseteq \AoneDC^*$. This indeed holds because $\AoneDC^*$ is closed and $\AnDC^* \subseteq \AoneDC^*$ for any $n$. The last fact is shown in Appendix \ref{app:WZ2}.

Now we turn our attention to proving
Theorems \ref{th:A1} and \ref{th:A1P}. Towards this goal we shall first derive in Sec. \ref{sec:funda}, a fundamental principle based on typicality that (upon appropriate generalization, as we shall see in \cite{PartII}) governs the inner bound of achievable region arising in any source coding problem with noncooperating encoders and one decoder. Based on this principle,
the proofs of Theorems \ref{th:A1} and \ref{th:A1P} will then be presented in Secs. \ref{sec:joint} and \ref{sec:partial}, respectively.

\Section{Fundamental Principle}
\label{sec:funda}
\subsection{Statement}

\begin{theorem}
\label{th:funda}
Let $(Y_1,Y_2,Z_1,Z_2)$ be four random variables following joint probability distribution $p'(y_1,y_2) q'_{1}(z_1|y_1) q'_2(z_2|y_2)$ and draw $\{(Y_{1}(k),Y_{2}(k))\}$ i.i.d. $\sim p'(y_1,y_2)$. 
Then for any rate pair $(R'_1,R'_2)$ such that
\bea
\label{eq:R1def}
I(Y_1;Z_1|Z_2) &\le& R'_1\\
\label{eq:R2def}
I(Y_2;Z_2|Z_1) &\le& R'_2\\
\label{eq:R12def}
I(Y_1,Y_2;Z_1,Z_2) &\le& R'_1+R'_2
\eea 
and any $\e'\rightarrow 0$,
there exists a sequence of mapping triplets $$(f_1:\Ycal_1^{n'} \rightarrow \Ucal_1,f_2:\Ycal_2^{n'} \rightarrow \Ucal_2,g:\Ucal_1\times\Ucal_2 \rightarrow \Zcal_1^{n'}\times \Zcal_2^{n'})$$ for some sequence of alphabet pairs $(\Ucal_1,\Ucal_2)$ (and some $n'\rightarrow\infty$) such that 
\bea
\label{eq:rate1}
\frac{1}{n'} \log |\Ucal_1| &\le& R'_1 + \e''\\
\label{eq:rate2}
\frac{1}{n'} \log |\Ucal_2| &\le& R'_2 + \e''\\
\label{eq:prob}
\Pr\{\Ecal\} &\le& \e''
\eea
where 
$$\Ecal =\{(Y_1^{n'},Y_2^{n'},
\Zhat_1^{n'},\Zhat_2^{n'})\notin \Tenn(Y_1,Y_2,Z_1,Z_2)\},$$
$(\Zhat_1^{n'},\Zhat_2^{n'})=g(f_1(Y_1^{n'}),f_2(Y_2^{n'}))$ and $\e''\rightarrow 0$.
\end{theorem}

Note that $Z_1\rightarrow Y_1 \rightarrow Y_2 \rightarrow Z_2$ is required to form Markov chain for Theorem \ref{th:funda} to apply.
Further, due to strong law of large numbers, the sequence $\{(Y_1^{n'},Y_2^{n'})\}$ in the above can be replaced without loss of generality by any sequence $\{(\hat{Y}_1^{n'},\hat{Y}_2^{n'})\}$ such that $\Pr\{(\hat{Y}_1^{n'},\hat{Y}_2^{n'})
\notin \Tenn(Y_1,Y_2)\}\le \e'_1$ where $\e'_1\rightarrow 0$ as $\e'\rightarrow 0$. Such substitutions are standard and will sometimes be carried out without explicit mention.
Also
note that, as $\e'\rightarrow 0$, $n'\rightarrow\infty$ through values $n'>n'_0(\e')$ for some appropriate $n'_0(\cdot)$. 

Theorem \ref{th:funda} roughly states the following. Using a sequence of codes $(f_1,f_2,g)$ (of sufficiently large length $n'$), one can achieve any rate pair $(R'_1,R'_2)$ satisfying conditions (\ref{eq:R1def})--(\ref{eq:R12def}) such that the estimate $(\Zhat_1^{n'},\Zhat_2^{n'})=g(f_1(Y_1^{n'}),f_2(Y_2^{n'}))$ of $(Z_1^{n'},Z_2^{n'})$ based on the encoding $(f_1(Y_1^{n'}),f_2(Y_2^{n'}))$ 
is jointly typical with $(Y_1^{n'},Y_2^{n'})$ with high probability. 
Further, Theorem \ref{th:funda} includes the direct statement of Slepian--Wolf theorem as a special case. To see this, set
$(Z_1,Z_2)=(Y_1,Y_2)$ and note that the left hand sides of (\ref{eq:R1def})--(\ref{eq:R12def}) reduce to $H(Y_1|Y_2)$, $H(Y_2|Y_1)$ and $H(Y_1,Y_2)$, respectively. Now, for any rate pair $(R'_1,R'_2)$ satisfying (\ref{eq:R1def})--(\ref{eq:R12def}), there exists a code $(f_1,f_2,g)$ such that (\ref{eq:rate1}) and (\ref{eq:rate2}) hold and 
$\Pr\{(\Yhat_1^{n'},\Yhat_2^{n'})\ne (Y_1^{n'},Y_2^{n'})\}$ is small due to (\ref{eq:prob}). This is the desired direct statement.

 Notice that Theorem \ref{th:funda} makes no reference to any distortion criterion. Yet we shall see that Theorem \ref{th:funda} plays a central role in proving the inner bounds in Theorems \ref{th:A1} and \ref{th:A1P}, 
each of which solves a problem where reconstruction is based on a distortion criterion. To give an insight into how Theorem \ref{th:funda} can be applied to a problem with a distortion criterion,
let us sketch the proof of the inner bound arising in Theorem \ref{th:A1} (which we subsequently formalize in Sec.~\ref{sec:2In}). To avoid technicality,
let us show only $\AD \supseteq \AD^*$ for the time being.
First of all, for any triplet $(R_1,R_2,D)\in \AD^* = \bigcup_{n=1}^\infty \AnD^*$, (\ref{eq:R1})--(\ref{eq:D12}) hold for some $n$. Now recalling that 
$Z_1\rightarrow X_1^n \rightarrow X_2^n \rightarrow Z_2$ is Markov chain and
identifying $(Y_1,Y_2)=(X_1^n,X_2^n)$, note that
(\ref{eq:R1def})--(\ref{eq:R12def}) hold for $R_1' = nR_1$ and $R_2'=nR_2$. Therefore, by Theorem \ref{th:funda}, for any $\e \rightarrow 0$,
(\ref{eq:rate1})--(\ref{eq:prob}) hold for some sequence of mapping triplets $(f_1,f_2,g)$. Note that (\ref{eq:rate1}) are (\ref{eq:rate2}) are identical to conditions (\ref{eq:AR1}) and (\ref{eq:AR2}) with $(\Ucal_1,\Ucal_2, nn')$ now playing the role of $(\Zcal_1,\Zcal_2,n)$. We are only left to see that the distortion given by the left hand side of (\ref{eq:ALC}), with $nn'$ now in place of $n$, is close to $D$. For this, denote  $$(\Xhat_1^{nn'},\Xhat_2^{nn'}) =\psi'(\Zhat_1^{n'},\Zhat_2^{n'}) = (\psi(\Zhat_1(1),\Zhat_2(1)),\psi(\Zhat_1(2),\Zhat_2(2)),
...,\psi(\Zhat_1(n'),\Zhat_2(n')))$$
and
observe that 
$(X_1^{nn'},X_2^{nn'})$ is jointly typical with $\psi'(\Zhat_1^{n'},\Zhat_2^{n'})$ with
high probability due to (\ref{eq:prob}). Consequently, in view of (\ref{eq:D12}), the distortion requirement is met as desired.

Next we prove Theorem \ref{th:funda} in three steps: The first two steps, given in Secs.~\ref{sec:step1} and \ref{sec:step2}, involve the derivation of two specializations, Lemmas \ref{le:adi1} and \ref{le:adi2}, respectively. The third step, given in Sec.~\ref{sec:step3}, completes the proof by combining the above lemmas. We shall see that Lemmas \ref{le:adi1} and \ref{le:adi2} capture the essence of the typicality arguments of Shannon \cite{Shannon} and Wyner-Ziv \cite{WZ}, respectively. 

\subsection{Derivation of Lemma \ref{le:adi1}}
\label{sec:step1}

\begin{lemma} 
\label{le:adi1}
Let $(Y,Z)\sim p'(y)q'(z|y)$ and draw $\{Y(k)\}$ i.i.d. $\sim p'(y)$.
Then for any rate $R'$ such that
\be
\label{eq:Rdef'}
I(Y;Z)\le R'
\ee
and any $\e'\rightarrow 0$,
there exists a sequence of 
mapping pairs
$$(f:\Ycal^{n'} \rightarrow \Ucal,g:\Ucal \rightarrow \Zcal^{n'})$$ for some sequence of alphabets $\Ucal$ (and some $n'\rightarrow\infty$) 
such that
\bea
\label{eq:ShRe}
\frac{1}{n'} \log |\Ucal| &\le& R' + \epsilon''\\
\label{eq:ShPr}
\Pr\{(Y^{n'},\Zhat^{n'}) \notin \Tenn(Y,Z)\} &\le& \epsilon''
\eea
where 
$\Zhat^{n'} = g(f(Y^{n'}))$
and $\e''\rightarrow 0$.
\end{lemma}

The direct statement of Shannon's rate-distortion theorem follows from Lemma \ref{le:adi1} in a straightforward manner. 
Following gives an outline.
Consider encoding of $Y\sim p'(y)$ under bounded distortion criterion $d$. Further, consider all conditional distributions $q'(z|y)$ such that $(Y,Z)\sim p'(x)q'(z|x)$ satisfies $\E d(Y,Z) \le D$. By Lemma \ref{le:adi1}, one can construct block codes $(f,g)$ of $Y^{n'}$ that achieve rates close to $I(Y;Z)$, while the decoded sequence $\Zhat^{n'}$ remains jointly typical with $Y^{n'}$ with high probability. Hence, reconstructing $\Yhat^{n'} = \Zhat^{n'}$, the distortion $\frac{1}{n'} \E d_{n'}(Y^{n'},\Yhat^{n'})$ remains close to $D$. This is Shannon's direct theorem \cite{Shannon}. The above argument will be extended in Sec.~\ref{sec:S}. 

The above analysis demonstrates that the essence of Shannon's 
direct theorem is abstracted in Lemma \ref{le:adi1} which makes no reference to the distortion criterion $d$. In fact, Lemma \ref{le:adi1} will play a significant role in the development of a unified source coding theory. Proof of Lemma \ref{le:adi1} is based on a random coding argument which borrows from the classical derivation of Shannon's direct theorem \cite{Cover}. To proceed, we need intermediate results given in Lemmas~ \ref{le:math} and \ref{le:typ}. 

\begin{lemma}
\label{le:math}
$1- x \le e^{-x}$ for $x\ge 0$.
\end{lemma}

{\bf \em Proof:} Denote $f(x) = e^{-x} + x -1$. Differentiating, we obtain $f'(x) = -e^{-x} +1 \ge 0$ for $x\ge 0$. Hence $f(x) \ge f(0) =0$ for $x\ge 0$ and the result follows. \hfill $\Box$

\begin{lemma}
\label{le:typ} \cite[Lemma 13.6.1]{Cover}
Let $(Y,Z)\sim p'(y)q'(z|y)$ and draw $\{\Zhat(k)\}$ i.i.d. $\sim p'_Z(z) =\sum_{y\in\Ycal} p'(y)q'(z|y)$. Then
\be
\label{eq:Pbnd1}
2^{-n' (I(Y;Z)+\e'_1) }\le \Pr\{(y^{n'},\Zhat^{n'})\in \Tenn(Y,Z) \} \le 2^{-n' (I(Y;Z)-\e'_1) }
\ee
for any $y^{n'} \in \Tenn(Y)$, where 
$\epsilon'_1 \rightarrow 0$ as $\epsilon' \rightarrow 0$ (and $n' \rightarrow \infty$, appropriately).
\end{lemma}

{\bf \em Proof of Lemma \ref{le:adi1}:} Pick an integer $K$ (a specific choice will be made later), draw each
$\Zhat^{n'}(1),\Zhat^{n'}(2),...,\Zhat^{n'}(K)$ i.i.d. $\sim \prod_{k=1}^{n'} p'_Z(z_k)$ (where $p'_Z(z) =\sum_{z\in\Zcal} p'(y)q'(z|y)$) independent of $Y^{n'}$, and define a random mapping $F$ taking values in the set of mappings $\Ycal ^{n'}\rightarrow \Zcal^{n'}$ in the following manner:
For any $y^{n'}\in \Ycal^{n'}$, if there exists $\Zhat^{n'}(i)$ such that $(y^{n'}, \Zhat^{n'}(i)) \in \Tenn(Y,Z)$, then set $F(y^{n'}) = \Zhat^{n'}(i)$ (if there are more than one such $\Zhat^{n'}(i)$, pick any one), otherwise assign $F(y^{n'})$ to arbitrary $\Zhat^{n'}(i)$.  Of course, each 
mapping $f$, such that $\Pr\{F=f\} >0$,
takes at most $K$ values, i.e., for any such $f$, we have
\be
\label{eq:Udef}
|\Ucal| \le K
\ee
where $\Ucal$ denotes the alphabet of $f(Y^{n'})$.
Also,  
for each ${y^{n'}\in \Tenn(Y)}$, we have 
\bea
\label{eq:E1''}
\Pr\{(y^{n'},F(y^{n'})) \notin \Tenn(Y,Z)\} &=& 
\prod_{i=1}^K \left(1- \Pr\{(y^{n'},\Zhat^{n'}(i)) \notin \Tenn(Y,Z)\}
\right)\\
\label{eq:aaa1}
&\le& \left(
1-
2^{-n' (I(Y;Z)+\e'_1) }
\right)^K\\
\label{eq:aaa2}
&\le & \exp\left(-2^{-n' (I(Y;Z)+\e'_1)}K\right),
\eea
where 
(\ref{eq:aaa1}) follows by 
Lemma \ref{le:typ} (note $\e'_1 \rightarrow 0$ as $\e' \rightarrow 0$), and
(\ref{eq:aaa2}) follows by using Lemma \ref{le:math} and setting $x= 2^{-n' (I(Y;Z)+\e'_1)}$.
Further, since $$\Pr\{(Y^{n'},F(Y^{n'})) \in \Tenn(Y,Z)| Y^{n'} \notin \Tenn(Y)\} =0,$$ we have
\bea
\label{eq:aa0}
&&
\!\!\!\!\!\!\!\!\!\!
\!\!\!\!\!\!\!
\Pr\{(Y^{n'},F(Y^{n'})) \notin \Tenn(Y,Z)\}\\
\nonumber
& =& 1-\Pr\{Y^{n'} \in \Tenn(Y)\} 
\Pr\{(Y^{n'},F(Y^{n'})) \in \Tenn(Y,Z)|Y^{n'} \in \Tenn(Y)\}\\
\label{eq:aa2}
& \le&  1- (1-\e')\left[1 - \exp\left(-2^{-n' (I(Y;Z)+\e'_1)}K\right) \right] 
\\
\label{eq:aa3}
& \le& \e' + \exp\left(-2^{-n' (I(Y;Z)+\e'_1)}K\right) = \e'_2,
\eea
say,
where (\ref{eq:aa2}) follows due to strong law of large numbers and by (\ref{eq:aaa2}). Now substitute the random mapping $F$ in (\ref{eq:aa0}) by some mapping value $f$ such that $\Pr\{F=f\} >0$ and upon substitution 
(\ref{eq:aa3}) still holds. Denoting the alphabet of $f(Y^{n'})$ by $\Ucal$ as before, of course, (\ref{eq:Udef}) holds.
Next choose $K$ (while adjusting $n'$, if necessary) such that
\be
\label{eq:Lchoose}
I(Y;Z) +2\e'_1\le \frac{1}{n'} \log K \le I(Y;Z) +3\e'_1
\ee
and, in (\ref{eq:aa3}), note that $\e'_2\rightarrow 0$ as $\e'\rightarrow 0$ by the lower bound in (\ref{eq:Lchoose})
and write
\be
\label{eq:ech}
\e'' = \max\left\{3\e'_1, \e'_2 \right\}.
\ee
Further, using (\ref{eq:Rdef'}), (\ref{eq:Udef}) and (\ref{eq:ech}) in the upper bound in (\ref{eq:Lchoose}), we obtain (\ref{eq:ShRe}).  
Finally, using (\ref{eq:ech}) in (\ref{eq:aa3}) (with $f$ now in place of $F$), (\ref{eq:ShPr}) follows. \hfill$\Box$  

\subsection{Derivation of Lemma \ref{le:adi2}}
\label{sec:step2}

\begin{lemma}
\label{le:adi2}
Let $(Y_1,Y_2,Z_1)\sim p'(y_1,y_2)q'_1(z_1|y_1)$ and draw $\{(Y_{1}(k),Y_{2}(k))\}$ i.i.d. $\sim p'(y_1,y_2)$.
Then for 
any rate $R'_1$ such that
\be
\label{eq:Rdef''}
I(Y_1;Z_1|Y_2)\le R'_1
\ee
and any $\e'\rightarrow 0$,
there exists a sequence of 
mapping pairs
$$(f_1:\Ycal_1^{n'} \rightarrow \Ucal_1,g:\Ucal_1\times \Ycal_2^{n'} \rightarrow \Zcal_1^{n'})$$
for some sequence of alphabets $\Ucal_1$ (and some $n'\rightarrow\infty$) 
such that
\bea
\label{eq:ShRe'}
\frac{1}{n'} \log |\Ucal_1| &\le& R'_1 + \epsilon''\\
\label{eq:ShPr'}
\Pr\{(Y_1^{n'},Y_2^{n'},\Zhat_1^{n'}) \notin \Tenn(Y_1,Y_2,Z_1)\} &\le& \epsilon''
\eea
where 
$\Zhat_1^{n'} = g(f_1(Y_1^{n'}),Y_2^{n'})$
 and $\e''\rightarrow 0$.
\end{lemma}

Wyner-Ziv's direct theorem follows from Lemma \ref{le:adi2} in the similar manner as Shannon's direct theorem followed from Lemma \ref{le:adi1}. An outline is given in the following.
Consider encoding of $Y_1$ under distortion criterion $d$ assuming complete availability of side information $Y_2$ at the decoder.
Further,
consider all conditional distributions $q'_1(z_1|y_1)$ such that $(Y_1,Y_2,Z_1)\sim p'(y_1,y_2)q'_1(z_1|y_1)$ satisfies $\E d(Y_1,\psi(Z_2,Y_2)) \le D$ for some mapping $\psi$.
By Lemma \ref{le:adi2}, one can construct complete side information codes $(f_1,g)$ that achieve rates $R_1'$ close to $I(Y_1;Z_1|Y_2)$, while the decoded sequence $\Zhat_1^{n'}$ is jointly typical with $(Y_1^{n'},Y_2^{n'})$ with high probability. Hence, $Y_1^{n'}$ coupled with the reconstruction
$$\Yhat_1^{n'} = (\psi(\Zhat_1(1),Y_2(1)),\psi(\Zhat_1(2),Y_2(2)),...,
\psi(\Zhat_1(n'),Y_2(n')))$$
gives a jointly typical sequence of $(Y_1,\psi(Z_1,Y_2))$ with high probability, which ensures that
the distortion $\frac{1}{n'} \E d_{n'}(Y^{n'},\Yhat^{n'})$ is close to $D$. This is Wyner-Ziv's direct theorem \cite{WZ}.

The above analysis demonstrates that the essence of Wyner-Ziv's 
direct theorem is abstracted in Lemma \ref{le:adi2} which makes no reference to the distortion criterion $d$. In fact, Lemma \ref{le:adi2} will play a significant role in the development of a unified source coding theory. Proof of Lemma \ref{le:adi2} is based on a random coding argument which borrows from the classical derivation of Wyner-Ziv's direct theorem \cite{Cover}. 
To proceed,
we need additional intermediate results given in Lemmas~ \ref{le:math'} and \ref{le:Mar}. 

\begin{lemma}
\label{le:math'}
$(1- x)^k \ge 1-kx$ for $x\in[0,1]$ and $k>0$.
\end{lemma}

{\bf \em Proof:} Denote $f(x) = (1-x)^k - (1-kx)$. Differentiating with respect to $x$, we obtain $f'(x) = {-k(1-x)}^{k-1} + k \ge 0$ for $x\in [0,1]$. Hence $f(x) \ge f(0) =0$ for $x\in [0,1]$, and the result follows. \hfill $\Box$

\begin{lemma}
\label{le:Mar} 
\cite[Lemma 14.8.1]{Cover}
Let $(Y_1,Y_2,Z_1)\sim p'(y_1,y_2)q'_1(z_1|y_1)$ and the sequence 
of triplets 
$\{(\hat{Y}_1(k),\hat{Y}_2(k),\hat{Z}_1(k))\}$
be such that, for any $\e'\rightarrow 0$ (and an appropriate $n'\rightarrow \infty$), 
\beann
\Pr\{(\hat{Y}_1^{n'},\hat{Y}_2^{n'})\notin\Tenn(Y_1,Y_2)\}&\le& \e'_1\\ 
\Pr\{\hat{Y}_1^{n'},\hat{Z}_1^{n'})\notin\Tenn(Y_1,Z_1)\} &\le& \e'_1
\eeann 
for some $\e'_1\rightarrow 0$. Then $$\Pr\{(\hat{Y}_1^{n'},\hat{Y}_2^{n'},\hat{Z}_1^{n'})
\notin\Tenn(Y_1,Y_2,Z_1)\}\le \e'_2$$ 
for some $\e'_2$ such that $\e'_2\rightarrow 0$ as $\e'\rightarrow 0$. 
\end{lemma}

The above is a rewording of the so-called Markov lemma given in \cite{Cover}.

{\bf \em Proof of Lemma \ref{le:adi2}:} Pick integers $K_1$  and $K_2$ and draw each
$\Zhat_1^{n'}(1),\Zhat_1^{n'}(2),...,\Zhat_1^{n'}(K_1)$ i.i.d. $\sim \prod_{k=1}^{n'} p_{Z_1}(z_{1}(k))$ independent of $(Y_1^{n'},Y_2^{n'})$, where $$
p_{Z_1}(z_{1}) = \sum_{(y_1,y_2)\in \Ycal_1\times \Ycal_2} p'(y_1,y_2)q'_1(z_1|y_1).$$ Next randomize encoder mapping $f_{1}$ by
defining a random mapping 
$F_{1}$ over the set of mappings $\Ycal_1 ^{n'}\rightarrow I_{K_2}$ in a two-step manner. First define random mapping $\Fhat_{1}$ over the set of mappings $\Ycal_1 ^{n'}\rightarrow I_{K_1}$ as follows:
For any $y_1^{n'}\in \Ycal_1^{n'}$, if there exists $\Zhat_1^{n'}(i)$ such that $(y_1^{n'}, \Zhat_1^{n'}(i)) \in \Tenn(Y_1,Z_1)$, then set $\Fhat_{1}(y_1^{n'}) = i$ (if there are more than one such $\Zhat^{n'}(i)$, pick any one and proceed); if there is no such $\Zhat_1^{n'}(i)$, 
assign $\Fhat_{1}(y^{n'})$ to arbitrary $i\in I_{K_1}$. Then defining a random variable (mapping) $T$ such that $T$ is independent of all preceding random variables and uniformly distributed over the set of mappings $I_{K_1}\rightarrow I_{K_2}$,
set $F_{1}(y^{n'}) = T(\Fhat_{1}(y_1^{n'}))$.  
Clearly, for each 
$f_{1}$ such that $\Pr\{F_{1}=f_{1}\} >0$, the alphabet $\Ucal_1$ of 
$f_1(Y_1^{n'})$ is a subset of $I_{K_2}$, i.e., 
\be
\label{eq:Udef'}
|\Ucal_1| \le K_2.
\ee
Correspondingly, randomize decoder mapping $g$ by the
random mapping 
$G$ defined over $I_{K_2}\times 
\Ycal_2^{n'}\rightarrow \Zcal_1^{n'}$ in the following manner:
For any $j\in I_{K_2}$, if there exists unique $\Zhat_1^{n'}(i)$ such that $(y_2^{n'}, \Zhat_1^{n'}(i)) \in \Tenn(Y_2,Z_1)$ and $T(i) =j$, then
set $G(j,y_2^{n'}) = \Zhat^{n'}_1(i)$; if there is no such $\Zhat_1^{n'}(i)$ or there are more than one, 
assign $G(j,y_2^{n'})$ to arbitrary $\Zhat^{n'}_1(i)$.

Now note that the probability that $(Y_1^{n'},Y_2^{n'})$ are not jointly typical is bounded by
\be
\label{eq:E0}
\Pr\{\Ecal_0\} = \Pr\{(Y_1^{n'},Y_2^{n'})\notin \Tenn(Y_1,Y_2)\}\le \e'
\ee
for sufficiently large $n'$. Next, the probability that $Y_2^{n'}$ is not jointly typical with $\Fhat_{1}(Y_1^{n'})$
(i.e., any of the $\Zhat_1^{n'}(i)$'s) is given by (recalling the steps (\ref{eq:E1''}) through (\ref{eq:aa3}))
\be
\label{eq:E1} 
\Pr\{\Ecal_1\} 
= \Pr\{(Y_1^{n'},\Fhat_{1}(Y_1^{n'})) \notin \Tenn(Y_1,Z_1)\}
\le 
\exp\left(-2^{-n' (I(Y_1;Z_1)+\e'_1)}K_1\right) +\e'
\ee
where $\e'_1\rightarrow 0$ as $\e'\rightarrow 0$. We shall make $\Pr\{\Ecal_1\}$ small 
by
choosing $K_1$ (while adjusting $n'$, if necessary) such that
\be
\label{eq:Lchoose1}
I(Y_1;Z_1) + 2\e'_1 \le \frac{1}{n'} \log K_1 \le I(Y_1;Z_1) + 3\e'_1.
\ee
Using the lower bound in (\ref{eq:E1}),
we have
\be 
\label{eq:E1'}
\Pr\{\Ecal_1\} 
\le \exp\left(-2^{n'\e'_1}\right)+\e'
\ee
which approaches zero as $\e'\rightarrow 0$ (and $n'\rightarrow\infty$, appropriately). Further, noting 
(\ref{eq:E0}) and (\ref{eq:E1'}) and applying Lemma \ref{le:Mar} (Markov lemma),  
the probability that $(Y_1^{n'},Y_2^{n'})$ is not jointly typical with $\Fhat_{1}(Y_1^{n'}))$
is bounded by 
\be
\label{eq:E2} 
\Pr\{\Ecal_2\} 
= \{(Y_1^{n'},Y_2^{n'},\Fhat_{1}(Y_1^{n'})) \notin \Tenn(Y_1,Y_2,Z_1)\}
\le \e'_2
\ee
where $\e'_2\rightarrow 0$ as $\e'\rightarrow 0$ (and $n'\rightarrow \infty$, appropriately). 

Next we bound the probability that there are more than one $i$'s such that $T(i) = F_{1}(Y_1^{n'})$ and
$(Y_2^{n'},\Zhat_1^{n'}(i)))$ are jointly typical. 
Define random variables $T'_j = T^{-1}(j)$, $j\in I_{K_2}$. 
Each $T'_j$ takes values $I\subseteq I_{K_1}$ and follows identical distribution because $T$ is uniformly distributed over all mappings $I_{K_1} \rightarrow I_{K_2}$. 
Now, for any $\e'_3>0$, make $n'$, $K_1$ and $K_2$ sufficiently large, keeping $K_1/K_2$ constant (e.g., by taking $K_1$ and $K_2$ in tandem  through multiples), such that 
\be
\label{eq:K'}
\E \left[ 
\left|{|T'_j|}^2 -\frac{K_1^2}{K_2^2}\right|\right] \le \e'_3
\ee
which is possible due to strong law of large numbers. Without loss of generality, 
choose $\e'_3$ such that $\e'_3\rightarrow 0$ as $\e'\rightarrow 0$.
Further, by Lemma \ref{le:typ}, we have, for
any ${y_2^{n'}\in \Tenn(Y_2)}$,
\be
\Pr\{(y_2^{n'},
\Zhat_1^{n'}(i)) \in \Tenn(Y_2,Z_1)\} \le 
2^{-n' (I(Y_2;Z_1)-\e'_4)} = x,
\ee
say, where $\e'_4\rightarrow 0$ as $\e'\rightarrow 0$ (and $n\rightarrow\infty$, appropriately). Moreover, for a 
given $T'(j) = I$, we have 
\bea
\nonumber
&&\Pr\left\{\left|\left\{i\in I:(y_2^{n'},
\Zhat_1^{n'}(i)) \in \Tenn(Y_2,Z_1)\right\}\right| > 1\right\}\\
\nonumber
&& \qquad \le 1 - (1-x)^{|I|} - |I| x (1-x)^{|I|-1}\\
\label{eq:a'aa2}
&& \qquad \le 1 - (1-|I|x)-|I|x(1- (|I|-1)x)\\
\label{eq:a'aa3}
&&\qquad \le |I|^2 x^2,
\eea
where (\ref{eq:a'aa2}) follows from Lemma \ref{le:math'}. 
Denoting 
\be
\label{eq:E3}
\Ecal_3 = \left\{\left|\left\{i\in T^{-1}(F_{1}(Y_1^{n'})):(Y_2^{n'},
\Zhat_1^{n'}(i)) \in \Tenn(Y_2,Z_1)\right\}\right| > 1
\right\},
\ee
replacing $|I|$ by the corresponding random variable $|T'(j)|$ in (\ref{eq:a'aa3})) and taking expectation, we obtain
\bea
\label{eq:E3'}
\Pr\{\Ecal_3\} \le \E[{|T'_j|}^2]x^2
&\le&
 \frac{K_1^2}{K_2^2}2^{-2n' (I(Y_2;Z_1)-\e'_4)} +\e'_3 2^{-2n' (I(Y_2;Z_1)-\e'_4)}
\eea
which follows from (\ref{eq:K'}) and by expanding $x=2^{-n' (I(Y_2;Z_1)-\e'_4)}$. We shall make $\Pr\{\Ecal_3\}$ small by choosing (while adjusting $n'$, if necessary)
\be
\label{eq:Lchoose2}
I(Y_2;Z_1) - 3\e'_4\le 
\frac{1}{n'} \log K_1 - \frac{1}{n'}\log K_2 \le I(Y_2;Z_1) -2\e'_4.
\ee
Specifically, we have
\be 
\label{eq:E3''}
\Pr\{\Ecal_3\} \le \exp\left(2^{-2n'\e'_4}\right)+ \e'_3 2^{-2n' (I(Y_2;Z_1)-\e'_4)}
\ee
which approaches zero as $\e'\rightarrow 0$ (and $n'\rightarrow \infty$, appropriately).

At this point, referring to (\ref{eq:E0}), (\ref{eq:E1}), (\ref{eq:E2}) and (\ref{eq:E3}), verify that the overall error event $\Ecal$ satisfies
\be
\label{eq:Ezl}
\Ecal =\{(Y_1^{n'},Y_2^{n'},G(F_{1}(Y_1^{n'}),Y_2^{n'})) \notin \Tenn(Y_1,Y_2,Z_1)\} \subseteq \Ecal_0\cup\Ecal_1\cup\Ecal_2\cup\Ecal_3
\ee
such that
\be
\label{eq:error}
\Pr\{\Ecal\}
\le \Pr\{\Ecal_0\}+ \Pr\{\Ecal_1\}+ \Pr\{\Ecal_2\}+\Pr\{\Ecal_3\} 
\le \e'_5
\ee
where $\e'_5 \rightarrow 0$ as $\e'\rightarrow 0$ (and $n'\rightarrow \infty$, appropriately) due to (\ref{eq:E0}), (\ref{eq:E1'}), (\ref{eq:E2}) and (\ref{eq:E3''}). Further, substitute the random mapping pair $(F_{1},G)$ in expression (\ref{eq:Ezl}) of $\Ecal$ by a mapping pair value $(f_{1},g)$ such that $\Pr\{(F_1,G) = (f_1,g)\} > 0$ and upon substitution (\ref{eq:error}) still holds. Denoting the alphabet of $f_1(Y_1^{n'})$ by $\Ucal_1$ as before, of course, (\ref{eq:Udef'}) holds.

At this point, using the upper bound in (\ref{eq:Lchoose1}) in the lower bound in (\ref{eq:Lchoose2}) and 
noting $I(Y_1;Z_1) - I(Y_2;Z_1) = I(Y_1;Z_1|Y_2)$ (because $Z_1 \rightarrow Y_1 \rightarrow Y_2$ is Markov chain), 
we have
\be
\label{eq:Lchoose3}
\frac{1}{n'}\log K_2 \le I(Y_1;Z_1|Y_2) + 3\e'_1 + 3\e'_4.
\ee
Now set 
\be
\label{eq:lopa}
\e'' = \max\left\{3\e'_1+3\e'_4, \e'_5\right\}
\ee
(of course, $\e'' \rightarrow 0$ as $\e'\rightarrow 0$). Using (\ref{eq:Rdef''}), (\ref{eq:Udef'}) and (\ref{eq:lopa}) in (\ref{eq:Lchoose3}), we obtain (\ref{eq:ShRe'}).  
Finally, using (\ref{eq:lopa}) in (\ref{eq:error}) (with $(f,g)$ now in place of $(F,G)$), (\ref{eq:ShPr'}) 
 follows. 
\hfill $\Box$ 

\subsection{Completion of Proof of Theorem \ref{th:funda}}
\label{sec:step3}

First, 
note
\bea
\nonumber
 I(Y_1;Z_1|Z_2) &=& H(Z_1|Z_2)-H(Z_1|Y_1,Z_2)\\
\label{eq:chain5}
&=& H(Z_1|Z_2) - H(Z_1|Y_1)\\
  \label{eq:chain6}
  &\le& H(Z_1) - H(Z_1|Y_1)\\ 
\label{eq:chain7}
&=& I(Y_1;Z_1)
\eea
where (\ref{eq:chain5}) follows because $Z_1\rightarrow Y_1 \rightarrow Z_2$ is Markov chain and (\ref{eq:chain6}) follows because conditioning reduces entropy.
Further,  
apply the chain rule and note
\bea
\nonumber
I(Y_1,Y_2;Z_1,Z_2) &=& I(Y_1,Y_2;Z_1) + I(Y_1,Y_2; Z_2|Z1)\\
\label{eq:chain1}
&=& I(Y_1;Z_1) + I(Y_2;Z_1|Y_1) + I(Y_2;Z_2|Z_1) + I(Y_1;Z_2|Z_1,Y_2).
\eea
Since $Z_1 \rightarrow Y_1 \rightarrow Y_2 \rightarrow Z_2$ is Markov chain, we have
\bea
\label{eq:chain2}
 I(Y_2;Z_1|Y_1) &=& 0\\
\nonumber
 I(Y_1;Z_2|Z_1,Y_2) &=& H(Z_2|Z_1,Y_2) - H(Z_2|Z_1,Y_2,Y_1)\\
  \label{eq:chain3}
  &=& H(Z_2|Y_2) - H(Z_2|Y_2) = 0. 
\eea
Consequently,
we conclude
\be
\label{eq:chain4}
I(Y_1,Y_2;Z_1,Z_2) = I(Y_1;Z_1)  + I(Y_2;Z_2|Z_1)
\ee
using (\ref{eq:chain2}) and (\ref{eq:chain3}) in (\ref{eq:chain1}).

At this point, denote by $\Bcal^*$ the set of $(R_1',R_2')$ satisfying (\ref{eq:R1def})--(\ref{eq:R12def}). Clearly, $\Bcal^*$ is convex. Further, we conclude the following from (\ref{eq:chain7}) and (\ref{eq:chain4}): 1) Any $({R'_1}^{(0)},{R'_2}^{(0)})$, such that 
\bea
\label{eq:giri1}
{R'_1}^{(0)} &\ge& I(Y_1;Z_1)\\
\label{eq:giri2}
{R'_2}^{(0)} &\ge& I(Y_2;Z_2|Z_1),
\eea
belongs to $\Bcal^*$; 2)
Any $({R'_1}^{(1)},{R'_2}^{(1)})$, such that 
\bea
\label{eq:giri1'}
{R'_1}^{(1)} &\ge& I(Y_1;Z_1|Z_2)\\
\label{eq:giri2'}
{R'_2}^{(1)} &\ge& I(Y_2;Z_2),
\eea 
also belongs to $\Bcal^*$, by symmetry; 3) Any $(R_1',R_2')\in \Bcal^*$ can be expressed as a convex combination of $({R'_1}^{(0)},{R'_2}^{(0)})$ and $({R'_1}^{(1)},{R'_2}^{(1)})$ such that (\ref{eq:giri1})--(\ref{eq:giri2'}) hold.

Further, denote by $\Bcal$ the set
of rate pairs $(R_1',R'_2)$ such that for any sequence $\e' \rightarrow 0$ there exists mapping triplet $(f_1,f_2,g)$ satisfying (\ref{eq:rate1})--(\ref{eq:prob}). We claim that $\Bcal$ is convex.
To see this, note that
if each of two rate pairs $({R'_1}^{(0)},{R'_2}^{(0)})$ and $({R'_1}^{(1)},{R'_2}^{(1)})$ belongs to $\Bcal$, by a time sharing argument, so does any convex combination of such pairs. Clearly, to complete the proof of Theorem \ref{th:funda}, 
we need to show $\Bcal^*\subseteq \Bcal$. Towards this goal, we claim
that any $({R'_1}^{(0)},{R'_2}^{(0)})$, satisfying (\ref{eq:giri1}) and (\ref{eq:giri2}), belongs to $\Bcal$. To see this, note the following. Firstly, by Lemma \ref{le:adi1}, for any $\e'\rightarrow 0$, there exists a sequence of mapping pairs $(f_1: \Ycal_1^{n'} \rightarrow \Ucal_1, g_1:\Ucal_1\rightarrow \Zcal_1^{n'})$ such that    
\bea
\label{eq:ShRe1}
\frac{1}{n'} \log |\Ucal_1| &\le& {R'_1}^{(0)} + \epsilon'_1\\
\label{eq:ShPr1}
\Pr\{(Y_1^{n'},\Zhat_1^{n'}) \notin \Tenn(Y_1,Z_1)\} &\le& \epsilon'_1
\eea
where $\Zhat_1^{n'} = g_1(f_1(Y_1^{n'}))$ and $\e'_1 \rightarrow 0$ as $\e' \rightarrow 0$. Also, in view of (\ref{eq:ShPr1}), there exists, by Lemma \ref{le:adi2}, a sequence of mapping pairs $(f_2: \Ycal_2^{n'} \rightarrow \Ucal_2, g_2:\Ucal_2\times\Zcal_1^{n'}\rightarrow \Zcal_2^{n'})$ 
such that    
\bea
\label{eq:ShRe1'}
\frac{1}{n'} \log |\Ucal_2| &\le& {R'_2}^{(0)} + \epsilon'_2\\
\label{eq:ShPr1'}
\Pr\{(Y_2^{n'},\Zhat_2^{n'}, \Zhat_1^{n'}) \notin \Tenn(Y_2,Z_2,Z_1)\} &\le& \epsilon'_2
\eea
where $\Zhat_2^{n'} = g_2(f_2(Y_2^{n'}),\Zhat_1^{n'})= g_2(f_2(Y_2^{n'}),g_1(f_1(Y_1^{n'})))$ and $\e'_2 \rightarrow 0$ as $\e' \rightarrow 0$. Noting (\ref{eq:ShPr1}), (\ref{eq:ShPr1'}) and the fact that $Z_1\rightarrow Y_1 \rightarrow Y_2 \rightarrow Z_2$ is Markov chain, apply Lemma \ref{le:Mar} (Markov lemma) repeatedly to obtain
\be
\label{eq:ShPr2}
\Pr\{(Y_1^{n'},Y_2^{n'}, \Zhat_1^{n'}, \Zhat_2^{n'}) \notin \Tenn(Y_1,Y_2,Z_1,Z_2)\} \le \epsilon'_3
\ee
where $$(\Zhat_1^{n'}, \Zhat_2^{n'}) = g(f_1(Y_1^{n'}),f_2(Y_2^{n'}))
= (g_1(f_1(Y_1^{n'})),g_2(f_2(Y_2^{n'}),g_1(f_1(Y_1^{n'}))))$$ 
and
$\e'_3 \rightarrow 0$ as $\e' \rightarrow 0$. By (\ref{eq:ShRe1}), (\ref{eq:ShRe1'}) and (\ref{eq:ShPr2}), respectively, and setting $\e'' = \max\{\e'_1,\e'_2,\e'_3\}$, (\ref{eq:rate1})--(\ref{eq:prob}) hold. Hence $({R'_1}^{(0)},{R'_2}^{(0)})\in \Bcal$. By symmetry, any $({R'_1}^{(1)},{R'_2}^{(1)})$, satisfying (\ref{eq:giri1'}) and (\ref{eq:giri2'}), also belongs to $\Bcal$. 

Now, recalling that
any $({R'_1},{R'_2})\in \Bcal^*$ can be expressed as a convex combination of some $({R'_1}^{(0)},{R'_2}^{(0)})$ and $({R'_1}^{(1)},{R'_2}^{(1)})$ such that (\ref{eq:giri1})--(\ref{eq:giri2'}) hold and noting that such pairs satisfy $({R'_1}^{(0)},{R'_2}^{(0)})\in \Bcal$ and $({R'_1}^{(1)},{R'_2}^{(1)})\in \Bcal$ (by the argument given in the last paragraph), we have $({R'_1},{R'_2})\in \Bcal$ due to convexity of $\Bcal$. Hence $\Bcal^*\subseteq \Bcal$ and the proof is complete.
 \hfill$\Box$

\Section{Proof of Theorem \ref{th:A1}}
\label{sec:joint}
Now we turn to the proof of Theorem \ref{th:A1} which consists of 
two parts: The inner bound $\AD \supseteq \ADSol$ is shown in Sec. \ref{sec:2In} using the fundamental principle given in Theorem \ref{th:funda}. The outer bound $\AD \subseteq \ADSol$ is shown in Sec. \ref{sec:2Out} with the aid of Slepian-Wolf theorem \cite{SW}. 

\subsection{Inner Bound \boldmath{$\Acal_{\mbox{\tiny \bf D}} \supseteq \overline{\Acal_{\mbox{\tiny \bf D}}^*}$}} 
\label{sec:2In}

For any triplet $(R_1,R_2,D)\in \AD^* = \bigcup_{n=1}^\infty \AnD^*$,
 (\ref{eq:R1})--(\ref{eq:D12}) hold for some 
random variables $(Z_1,Z_2)$ such that $Z_1 \rightarrow X_1^n \rightarrow X_2^n \rightarrow Z_2$ is a Markov chain and for some mapping $\psi:\Zcal_1\times\Zcal_2 \rightarrow \Xcal_1^n\times\Xcal_2^n$. 
Referring to  Theorem \ref{th:funda}, note that conditions (\ref{eq:R1})--(\ref{eq:R12})
are same as conditions (\ref{eq:R1def})--(\ref{eq:R12def}) upon identifying
$(Y_1,Y_2,Z_1,Z_2)$ with $(X_1^n,X_2^n,Z_1,Z_2)$ and 
$(R'_1,R'_2)$ with $(nR_1,nR_2)$.
Consequently, by Theorem \ref{th:funda}, for any $\e' \rightarrow 0$, there exists a sequence of mapping triplets
$$(f_1:\Xcal_1^{nn'} \rightarrow \Ucal_1,f_2:\Xcal_2^{nn'} \rightarrow \Ucal_2,g:\Ucal_1\times\Ucal_2 \rightarrow \Zcal_1^{n'}\times \Zcal_2^{n'})$$ (for some $n'\rightarrow \infty$)
such that (\ref{eq:rate1})--(\ref{eq:prob}) hold. In other words, we respectively have (the first two conditions (\ref{eq:rate1}) and (\ref{eq:rate2}) are divided throughout by $n$)
\bea
\label{eq:rate1J}
\frac{1}{nn'} \log |\Ucal_1| &\le& R_1 + \e''/n\\
\label{eq:rate2J}
\frac{1}{nn'} \log |\Ucal_2| &\le& R_2 + \e''/n\\
\label{eq:probJ}
\Pr\{\Ecal\} &\le& \e''
\eea
where 
$${\cal E} = \{ (X_1^{nn'},X_2^{nn'},\Zhat_1^{n'},\Zhat_2^{n'}) \notin \Tenn(X_1^n,X_2^n,Z_1,Z_2)\},$$
$(\Zhat_1^{n'},\Zhat_2^{n'})=g(f_1(X_1^{nn'}),f_2(X_2^{nn'}))$ and $\e''\rightarrow 0$. 
Further, denote   
\beann
(\Xhat_1^{n}(j),\Xhat_2^{n}(j)) &=& \psi(\Zhat_1(j),\Zhat_2(j)),\quad 
1\le j\le n'\\
(\Xhat_1^{nn'},\Xhat_2^{nn'}) &=& \psi'(\Zhat_1^{n'},\Zhat_2^{n'}) 
=g'(f_1(X_1^{nn'}),f_2(X_2^{nn'})).
\eeann
Note that $g':\Ucal_1\times\Ucal_2 \rightarrow \Xcal_1^{nn'}\times \Xcal_2^{nn'}$.
Due to (\ref{eq:probJ}), we have (adjusting $n'\rightarrow \infty$, if necessary)
\be
\label{eq:prob'}
\Pr\{\Ecal_1\} \le \e''_1
\ee
where
\[
\Ecal_1 = \{(X_1^{nn'},X_2^{nn'},\psi'(\Zhat_1^{n'},\Zhat_2^{n'}))\notin \Tenn(X_1^n,X_2^n,\psi(Z_1,Z_2))\}
\]
and 
$\e''_1\rightarrow 0$ as $\e'\rightarrow 0$. Also, due to (\ref{eq:D12}), we have
$$\frac{1}{nn'}d_{nn'}((x_1^{nn'},x_2^{nn'}),(\xhat_1^{nn'},\xhat_2^{nn'}))\le D +\e'\dmax$$ for any $(x_1^{nn'},x_2^{nn'},(\xhat_1^{nn'},\xhat_2^{nn'}))\in \Tenn(X_1^{n},X_2^{n},\psi(Z_1,Z_2))$. Hence  
we obtain
\bea
\nonumber
\frac{1}{nn'} 
\E \,d_{nn'}((X_1^{nn'},X_1^{nn'}),(\Xhat_1^{nn'},\Xhat_2^{nn'}))
&\le& (1- \Pr\{ {\cal E}_1\}) (D +\e'\dmax) +  \Pr\{ {\cal E}_1\} d_{\max}\\
\label{eq:xc3}
&\le& D +(\e' + \e''_1) d_{\max}
\eea
using (\ref{eq:prob'}). Now, for a particular $\e>0$,
choose
$\e'>0$ such that $\max\{\epsilon'',(\e' + \e''_1) d_{\max}\}\le \e$. Consequently, conditions (\ref{eq:rate1J}), (\ref{eq:rate2J}) and (\ref{eq:xc3}) give rise to conditions (\ref{eq:AR1})--(\ref{eq:ALC}), respectively, with $(\Ucal_1, \Ucal_2,g',nn')$ now playing the role of $(\Zcal_1, \Zcal_2,g,n)$. Hence $(R_1,R_2,D)\in \AD$. In other words, $\AD\supseteq\AD^*$. Since $\AD$ is closed, we have $\AD\supseteq\ADSol$ (noting $\ADSol$ is the smallest closed set with $\AD^*$ as a subset).
This completes the proof.
\hfill$\Box$

\subsection{Outer Bound \boldmath{$\Acal_{\mbox{\tiny \bf D}} \subseteq \overline{\Acal_{\mbox{\tiny \bf D}}^*}$}} 
\label{sec:2Out}

Recall that any triplet $(R_1,R_2,D)\in \AD$ if, for any $\e>0$, there exists joint distortion code
$$(f_{1}:\Xcal_1^n \rightarrow \Zcal_1,
f_{2}:\Xcal_2^n \rightarrow \Zcal_2,
g: \Zcal_1\times\Zcal_2 \rightarrow \Xcal_1^n\times \Xcal_2^n)$$
of sufficiently large length $n$
such that (\ref{eq:AR1})--(\ref{eq:ALC}) hold. For easy reference,
the above conditions are reproduced below:
\bea
\label{eq:AR1rep}
\frac{1}{n} \log |\Zcal_1| &\le& R_1 +\e\\
\label{eq:AR2rep}
\frac{1}{n} \log |\Zcal_2| &\le& R_2 +\e\\
\label{eq:ALCrep}
\frac{1}{n} \E d_{n}((X_1^{n},X_2^{n}), 
g(f_1(X_1^{n}),f_2(X_2^n)) )
&\le& D +\e.
\eea
As seen in \cite{WZ}, we can further 
encode 
$(Z_1,Z_2) =(f_1(X_1^n),f_2(X_2^n))$
using an interposed Slepian--Wolf code 
$$(f'_{1}:\Zcal_1^{n'} \rightarrow \Ucal_1,f'_{2}:\Zcal_2^{n'} \rightarrow \Ucal_2,
g': \Ucal_1\times\Ucal_2 \rightarrow \Zcal_1^{n'}\times \Zcal_2^{n'}).$$
Given $(R_1,R_2,D)\in \AD$, $\e$ and $(f_1,f_2,g)$, 
a rate pair $(R'_1,R'_2)$ is said to be achievable using interposed codes if for any $\e'>0$ there exists mapping triplet $(f'_1,f'_2,g')$ (of length $n'$) such that
\bea
\label{eq:ax1'a}
\frac{1}{n'} \log |\Ucal_1| &\le& R'_1 +\e'\\
\label{eq:ax1'b}
\frac{1}{n'} \log |\Ucal_2| &\le& R'_2 +\e'\\
\label{eq:ax2'}
\Pr\{(Z_1^{n'},Z_2^{n'})\ne g'(f'_1(Z_1^{n'}),f'_2(Z_2^{n'}))\} &\le & \e'.
\eea
In view of (\ref{eq:AR1rep}) and (\ref{eq:AR2rep}),
setting
each of $f'_1$, $f'_2$ and $g'$ to identity mapping (clearly, $n'=1$, $\Ucal_1=\Zcal_1$, $\Ucal_2 = \Zcal_2$), of course, (\ref{eq:ax1'a})--(\ref{eq:ax2'}) trivially hold for $(R'_1,R'_2) = (n(R_1+\e),n(R_2+\e))$ irrespective of $\e'$. Therefore,
we have
\bea
\label{eq:R1.}
\onen
H(f_1(X_1^n)|f_2(X_2^n)) &\le& R_1+\e \\
\label{eq:R2.}
\onen
H(f_2(X_2^n)|f_1(X_1^n)) &\le& R_2+\e\\
\label{eq:R12.}
\onen
H(f_1(X_1^n),f_2(X_2^n)) &\le& R_1 + R_2 + 2\e
\eea
by Slepian-Wolf theorem \cite{Cover}. 

At this point define for any $\e\ge 0$ and any integral $n\ge 1$ the set $\AnD^{*(\e)}$ of rate-distortion triplets $(R_1,R_2,D)$ such that there exist alphabets $\Zcal_1$ and $\Zcal_2$, conditional distributions $q_1(z_1|x_1^n)$ and $q_2(z_2|x_2^n)$, and mapping $\psi:\Zcal_1\times\Zcal_2 \rightarrow \Xcal_1^n\times\Xcal_2^n$, satisfying
\bea
\label{eq:R1x}
\onen
I(X_1^n;Z_1|Z_2) &\le& R_1+\e \\
\label{eq:R2x}
\onen
I(X_2^n;Z_2|Z_1) &\le& R_2+\e \\
\label{eq:R12x}
\onen
I(X_1^n,X_2^n;Z_1,Z_2) &\le& R_1 + R_2+2\e \\
\label{eq:D12x}
\onen d_{n}((X_1^n,X_2^n),\psi(Z_1,Z_2))&\le& D+\e
\eea
where 
$(X_1^n,X_2^n,Z_1,Z_2) \sim 
p_n(x_1^n,x_2^n)
q_{1}(z_1|x_1^n) q_{2}(z_2|x_2^n)$ (i.e., $Z_1 \rightarrow X_1^n \rightarrow X_2^n \rightarrow Z_2$ is a Markov chain)
and $p_n(x_1^n,x_2^n) =\prod_{k=1}^n p(x_{1}(k),x_{2}(k))$. 
Comparing (\ref{eq:R1x})--(\ref{eq:D12x}) with (\ref{eq:R1})--(\ref{eq:D12}), note that $\AnD^{*(0)}=\AnD^*$.
In addition, 
$\AnD^{*(\e_1)}\subseteq\AnD^{*(\e_2)}$ for $\e_1\le \e_2$.
Further, let $\AD^{*(\e)} = \bigcup_{n=1}^\infty \AnD^{*(\e)}$.
Of course, $\AD^{*(\e_1)}\subseteq\AD^{*(\e_2)}$ for $\e_1\le \e_2$. 
Hence, noting $\bigcap_{\e>0} \overline{\AD^{*(\e)}}$ is closed, we obtain
\be
\label{eq:lope}
\bigcap_{\e>0} \overline{\AD^{*(\e)}} = \overline{\AD^{*(0)}} = \overline{\AD^*}.
\ee
The second equality in (\ref{eq:lope}) holds because $\AD^{*(0)} = \bigcup_{n=1}^\infty \AnD^{*(0)}= \bigcup_{n=1}^\infty \AnD^*= \AD^*$. 

Finally, consider any $(R_1,R_2,D)\in \AD$. Recall that for any $\e>0$ there exists joint distortion code $(f_1,f_2,g)$ such that (\ref{eq:R1.})--(\ref{eq:R12.}) and (\ref{eq:ALCrep}) hold. Choosing $(Z_1,Z_2) = (f_1(X_1^n),f_2(X_2^n))$ (of course, $Z_1 \rightarrow X_1^n \rightarrow X_2^n \rightarrow Z_2$ is then a Markov chain) and $\psi=g$, note that the above four conditions coincide with (\ref{eq:R1x})--(\ref{eq:D12x}), respectively. Hence $(R_1,R_2,D)\in \AnD^{*(\e)}$. Consequently, we have $(R_1,R_2,D)\in \AD^{*(\e)}\subseteq \overline{\AD^{*(\e)}}$ for each $\e>0$. Hence, by (\ref{eq:lope}), we have $(R_1,R_2,D)\in \bigcap_{\e>0} \overline{\AD^{*(\e)}}=\overline{\AD^*}$. This completes the proof. \hfill$\Box$

\Section{Proof of Theorem \ref{th:A1P}}
\label{sec:partial}
The proof is similar to the proof of Theorem \ref{th:A1} and
consists of two parts again: The inner bound $\ADP \supseteq \ADPSol$ is shown in Sec. \ref{sec:2pIn} using Lemmas \ref{le:adi1} and \ref{le:adi2} (both specializing Theorem \ref{th:funda}). The outer bound $\ADP \subseteq \ADPSol$ is shown in Sec. \ref{sec:2pOut} using Shannon's lossless coding theorem \cite{ShanLL} and Slepian-Wolf theorem \cite{SW}. 

\subsection{Inner Bound \boldmath{$\Acal_{\mbox{\tiny \bf DP}} \supseteq \overline{\Acal_{\mbox{\tiny \bf DP}}^*}$}} 
\label{sec:2pIn}

Since $\ADP^*= \bigcup_{n=1}^\infty \AnDP^*$,
for any triplet $(R_1,R_2,D)\in \ADP^*$,
 (\ref{eq:R1P}), (\ref{eq:R2P}) and (\ref{eq:D1P}) hold for some 
$\psi:\Zcal_1\times\Zcal_2 \rightarrow \Xcal_1^n$ and random variables $(Z_1,Z_2)$ such that $Z_1 \rightarrow X_1^n \rightarrow X_2^n \rightarrow Z_2$ is a Markov chain. 
Referring to Lemma \ref{le:adi1}, note that (\ref{eq:R2P}) is same as (\ref{eq:Rdef'}) for the choice $(Y,Z)=(X^n,Z_2)$ and 
$R' = nR$.
Consequently, by Lemma \ref{le:adi1}), for any $\e' \rightarrow 0$, there exists a sequence of mapping pairs
$$(f_2:\Xcal_2^{nn'} \rightarrow \Ucal_2,g_2:\Ucal_2 \rightarrow \Zcal_2^{n'})$$ (for some $n'\rightarrow \infty$)
such that (\ref{eq:ShRe}) and (\ref{eq:ShPr}) hold. 
In other words, we respectively have ((\ref{eq:ShRe}) is divided throughout by $n$)
\bea
\label{eq:rate2P}
\frac{1}{nn'} \log |\Ucal_2| &\le& R_2 + \e''/n\\
\label{eq:prob2P}
\Pr\{(X_2^{nn'},\Zhat_2^{n'}) \notin \Tenn(X_2^n,Z_2)\} &\le& \e''
\eea
where 
$\Zhat_2^{n'}=g_2(f_2(X_2^{nn'}))$ and $\e''\rightarrow 0$ as $\e'\rightarrow 0$. 
Noting (\ref{eq:prob2P}) and the fact that $X_1^{n} \rightarrow X_2^{n} \rightarrow Z_2$ is a Markov chain, we have, by Lemma \ref{le:Mar},
\be
\label{eq:MarP}
\Pr\{(X_1^{nn'},X_2^{nn'},\Zhat_2^{n'}) \notin \Tenn(X_1^n,X_2^n,Z_2)\} \le \e'_1
\ee
where $\e'_1\rightarrow 0$ as $\e'\rightarrow 0$.
Further, 
referring to Lemma \ref{le:adi2}, note that (\ref{eq:R1P}) is same as (\ref{eq:Rdef''}), where $(X_1^n,Z_2,Z_1)$ (note $Z_1\rightarrow X_1^n \rightarrow Z_2$ is a Markov chain) 
now plays the role of $(Y_1,Y_2,Z_1)$ and 
$R'_1 = nR_1$. 
Consequently, by Lemma \ref{le:adi2}), for any $\e' \rightarrow 0$, there exists sequence of mapping pairs
$$(f_1:\Xcal_1^{nn'} \rightarrow \Ucal_1,g_1:\Ucal_1\times\Zcal_2^{n'} \rightarrow \Zcal_1^{n'})$$ (for some $n'\rightarrow \infty$)
such that
(\ref{eq:ShRe'}) and (\ref{eq:ShPr'}) hold.
In other words, we respectively have ((\ref{eq:ShRe'}) is divided throughout by $n$)
\bea
\label{eq:rate1P}
\frac{1}{nn'} \log |\Ucal_1| &\le& R_1 + \e''_1/n\\
\label{eq:prob1P}
\Pr\{(X_1^{nn'},\Zhat_2^{n'},\Zhat_1^{n'}) \notin \Tenn(X_2^n,Z_2,Z_1)\} &\le& \e''_1
\eea
where 
$\Zhat_1^{n'}=g_1(f_1(X_1^{nn'}),\Zhat_2^{n'})$ and $\e''_1\rightarrow 0$ as $\e'\rightarrow 0$. 
Further, denote   
\beann
\Xhat_1^{n}(j) &=& \psi(\Zhat_1(j),\Zhat_2(j)),\quad 
1\le j\le n'\\
\Xhat_1^{nn'} &=& \psi'(\Zhat_1^{n'},\Zhat_2^{n'}) 
=g(f_1(X_1^{nn'}),f_2(X_2^{nn'})).
\eeann
Clearly, $g:\Ucal_1\times\Ucal_2 \rightarrow \Xcal_1^{nn'}$.
Due to (\ref{eq:prob1P}), we have (adjusting $n'\rightarrow \infty$, if necessary)
\be
\label{eq:prob''}
\Pr\{\Ecal_1\} \le \e''_2
\ee
where
\[
\Ecal_1 = \{(X_1^{nn'},\psi'(\Zhat_1^{n'},\Zhat_2^{n'})\notin \Tenn(X_1^n,\psi(Z_1,Z_2))\}
\]
and 
$\e''_2\rightarrow 0$ as $\e'\rightarrow 0$.
Since
$$\frac{1}{nn'}d_{nn'}(x_1^{nn'},\xhat_1^{nn'})\le D +\e'\dmax$$ for any $(x_1^{nn'},\xhat_1^{nn'})\in \Tenn(X_1^{n},\psi(Z_1,Z_2))$ due to (\ref{eq:D1P}), 
we have  
\bea
\nonumber
\frac{1}{nn'} 
\E \,d_{nn'}(X_1^{nn'},\Xhat_1^{nn'})
&\le& (1- \Pr\{ {\cal E}_1\}) (D +\e'\dmax) +  \Pr\{ {\cal E}_1\} d_{\max}\\
\label{eq:xc3'}
&\le& D +(\e' + \e''_2) d_{\max}
\eea
due to (\ref{eq:prob''}). 

Now, for a particular $\e>0$, choose
$\e'$ such that $\max\{\epsilon'',\e''_1,(\e' + \e''_2) d_{\max}\}\le \e$. Hence conditions (\ref{eq:rate1P}), (\ref{eq:rate2P}) and (\ref{eq:xc3'})
give rise to conditions (\ref{eq:AR1}), (\ref{eq:AR2}) and (\ref{eq:ALC'}), where $(\Ucal_1,\Ucal_2,nn')$ now take the place of 
$(\Zcal_1,\Zcal_2,n)$. 
Hence $(R_1,R_2,D)\in \ADP$. 
In other words, $\ADP\supseteq\ADP^*$. Since $\ADP$ is closed, we have $\ADP\supseteq\ADPSol$ (noting $\ADPSol$ is the smallest closed set with $\ADP^*$ as a subset).
This completes the proof.
\hfill$\Box$

\subsection{Outer Bound \boldmath{$\Acal_{\mbox{\tiny \bf DP}} \subseteq \overline{\Acal^*_{\mbox{\tiny \bf DP}}}$}} 
\label{sec:2pOut}

Recall that any triplet $(R_1,R_2,D)\in \ADP$ if, for any $\e>0$, there exists partial side information code
$$(f_{1}:\Xcal_1^n \rightarrow \Zcal_1,
f_{2}:\Xcal_2^n \rightarrow \Zcal_2,
g: \Zcal_1\times\Zcal_2 \rightarrow \Xcal_1^n)$$
of some length $n$
such that (\ref{eq:AR1}), (\ref{eq:AR2}) and (\ref{eq:ALC'}) hold. 
The above conditions are reproduced below for easy reference:
\bea
\label{eq:AR1rep'}
\frac{1}{n} \log |\Zcal_1| &\le& R_1 +\e\\
\label{eq:AR2rep'}
\frac{1}{n} \log |\Zcal_2| &\le& R_2 +\e\\
\label{eq:ALCrep'}
\frac{1}{n} \E d_{n}(X_1^{n}, 
g(f_1(X_1^{n}),f_2(X_2^n)) )
&\le& D +\e.
\eea
We can further encode $Z_1=f_1(X_1^{n})$ with complete side information $Z_2=f_2(X_2^n)$ using Slepian-Wolf code 
$$(f'_{1}:\Zcal_1^{n'} \rightarrow \Ucal_1,g'_{1}:\Ucal_1\times\Zcal_2^{n'} \rightarrow \Zcal_1^{n'}).$$
Given $(R_1,R_2,D)\in \AD$, $\e$ and $(f_1,f_2,g)$, 
rate $R'_1$ is said to be achieved using interposed Slepian-Wolf codes if for any $\e'>0$ there exists $(f'_1,g'_1)$ (of length $n'$) such that
\bea
\label{eq:ei1}
\frac{1}{n'} \log |\Ucal_1| &\le& R'_1 +\e'\\
\label{eq:ei2}
\Pr\{Z_1^{n'}\ne g'_1(f'_1(Z_1^{n'}),Z_2^{n'})\} &\le & \e'.
\eea
In view of (\ref{eq:AR1rep'}),
setting $f'_1$ to identity mapping (clearly, $n'=1$, $\Ucal_1=\Zcal_1$) and choosing $g'_1(f'_1(Z_1^{n'}),Z_2^{n'})=f'_1(Z_1^{n'}) = Z_1^{n'}$, of course, (\ref{eq:ei1}) and (\ref{eq:ei2}) trivially hold for $R'_1 = n(R_1+\e)$ irrespective of $\e'$. Therefore,
we have
\be
\label{eq:R1.'}
\onen
H(f_1(X_1^n)|f_2(X_2^n)) \le R_1+\e 
\ee 
by Slepian-Wolf theorem \cite{Cover}. In the same manner, we can also encode $Z_2=f_2(X_2^n)$
using an interposed Shannon code 
$$(f'_{2}:\Zcal_2^{n_2} \rightarrow \Ucal_2,g'_{2}:\Ucal_2 \rightarrow \Zcal_2^{n_2}).$$ 
Again
given $(R_1,R_2,D)\in \AD$, $\e$ and $(f_1,f_2,g)$, 
rate $R'_2$ is said to be achieved using interposed Shannon codes if for any $\e'>0$ there exists $(f'_2,g'_2)$ (of length $n'$) such that
\bea
\label{eq:ei1'}
\frac{1}{n'} \log |\Ucal_2| &\le& R'_2 +\e'\\
\label{eq:ei2'}
\Pr\{Z_2^{n'}\ne g'_2(f'_2(Z_2^{n'}))\} &\le & \e'.
\eea
In view of (\ref{eq:AR2rep'}),
setting
each of $f'_2$ and $g'_2$ to identity mapping (clearly, $n'=1$, $\Ucal_2=\Zcal_2$), of course, (\ref{eq:ei1'}) and (\ref{eq:ei2'}) trivially hold for $R'_2 = n(R_2+\e)$ irrespective of $\e'$. Therefore,
we have
\be
\label{eq:R2.'}
\onen
H(f_2(X_2^n)) \le R_2+\e 
\ee 
by Shannon's lossless coding theorem \cite{Cover}.

At this point define for any $\e\ge 0$ and any integral $n\ge 1$ the set $\AnDP^{*(\e)}$ of rate-distortion triplets $(R_1,R_2,D)$ such that there exist alphabets $\Zcal_1$ and $\Zcal_2$, conditional distributions $q_1(z_1|x_1^n)$ and $q_2(z_2|x_2^n)$, and mapping $\psi:\Zcal_1\times\Zcal_2 \rightarrow \Xcal_1^n\times\Xcal_2^n$, satisfying
\bea
\label{eq:R1x'}
\onen
I(X_1^n;Z_1|Z_2) &\le& R_1+\e \\
\label{eq:R2x'}
\onen
I(X_2^n;Z_2) &\le& R_2+\e \\
\label{eq:D12x'}
\onen d_{n}(X_1^n,\psi(Z_1,Z_2))&\le& D+\e
\eea
where 
$(X_1^n,X_2^n,Z_1,Z_2) \sim 
p_n(x_1^n,x_2^n)
q_{1}(z_1|x_1^n) q_{2}(z_2|x_2^n)$ (i.e., $Z_1 \rightarrow X_1^n \rightarrow X_2^n \rightarrow Z_2$ is a Markov chain)
and $p_n(x_1^n,x_2^n) =\prod_{k=1}^n p(x_{1}(k),x_{2}(k))$. 
Comparing with (\ref{eq:R1P})-(\ref{eq:D1P}), note that $\AnDP^{*(0)}=\AnDP^*$.
In addition, 
$\AnDP^{*(\e_1)}\subseteq\AnDP^{*(\e_2)}$ for $\e_1\le \e_2$.
Further, let $\ADP^{*(\e)} = \bigcup_{n=1}^\infty \AnDP^{*(\e)}$.
Of course, $\ADP^{*(\e_1)}\subseteq\ADP^{*(\e_2)}$ for $\e_1\le \e_2$. 
Hence, noting $\bigcap_{\e>0} \overline{\ADP^{*(\e)}}$ is closed, we obtain
\be
\label{eq:lope'}
\bigcap_{\e>0} \overline{\ADP^{*(\e)}} = \overline{\ADP^{*(0)}} = \overline{\ADP^*}.
\ee
The second equality in (\ref{eq:lope'}) holds because $\ADP^{*(0)} = \bigcup_{n=1}^\infty \AnDP^{*(0)}= \bigcup_{n=1}^\infty \AnDP^*= \ADP^*$. 

Finally, consider any $(R_1,R_2,D)\in \ADP$. Recall that for any $\e>0$ there exists partial side information code $(f_1,f_2,g)$ such that (\ref{eq:R1.'}), (\ref{eq:R2.'}) and (\ref{eq:ALCrep'}) hold. Choosing $(Z_1,Z_2) = (f_1(X_1^n),f_2(X_2^n))$ (of course, $Z_1 \rightarrow X_1^n \rightarrow X_2^n \rightarrow Z_2$ is then a Markov chain) and $\psi=g$, note that the above three conditions coincide with (\ref{eq:R1x'})-(\ref{eq:D12x'}), respectively. Hence $(R_1,R_2,D)\in \AnDP^{*(\e)}$. 
Consequently, we have $(R_1,R_2,D)\in \ADP^{*(\e)}\subseteq \overline{\ADP^{*(\e)}}$ for each $\e>0$. Hence, by (\ref{eq:lope'}), we have $(R_1,R_2,D)\in \bigcap_{\e>0} \overline{\ADP^{*(\e)}}=\overline{\ADP^*}$.
This completes the proof. \hfill$\Box$

\Section{Known Coding Theorems}
\label{sec:known}
So far we have solved (in Theorems \ref{th:A1} and \ref{th:A1P}, respectively) the joint distortion and partial side information problems which hitherto remained open. More precisely, we have outlined a new methodology that expands older techniques and solves the above open problems. We now illustrate that our technique applies to the known problems equally well. In particular, we now outline based on our method proofs of four widely celebrated results, namely, Shannon's lossy coding theorem \cite{Shannon}, Wyner-Ziv theorem \cite{WZ}, side information theorem \cite{Wyner, AhlKor} and Berger-Yeung theorem \cite{BY}, in Secs. \ref{sec:Shan}--\ref{sec:BY}, respectively. 
First we compare our technique with its classical counterpart in Sec.~\ref{sec:comment}.

\subsection{Comparison with Classical Proofs}
\label{sec:comment}

Corresponding to each known problem under consideration, we shall define the achievable rate (rate-distortion) region $\Acal$. As seen in Sec.~\ref{sec:prob}, we shall also define rate (rate-distortion) regions $\Acal_n^*$ that have suitable $n$-th order information-theoretic descriptions and let $\Acal^* = \bigcup_{n=1}^\infty \Acal_n^*$. We subscript these sets by `S', `WZ', `SI' and `BY', respectively, to indicate the correspondence with the known theorems listed above. We know that each theorem describes the corresponding achievable region by a first order description of the form $\Acal=\Acal_1^*$ \cite{Cover,BY}. 

The classical proof of each of the above theorems consists of two steps: Proof of the direct theorem $\Acal\supseteq\Acal_1^*$
and proof of the converse $\Acal\subseteq\Acal_1^*$. 
Clubbing $n$ source symbols together to start with (where $n$ is arbitrarily chosen), one can of course show $\Acal\supseteq\Acal_n^*$ and $\Acal\subseteq\Acal_n^*$ using the same argument, thus in effect showing (since $\Acal_1^*$ is closed) $$\Acal=\AcalSol = \Acal^*=\Acal^*_1.$$ In fact, we have discovered (further details can be found in \cite{PartII}) that the coding theorems corresponding to a broad class of problems (including, of course, the above known results as well as Theorems \ref{th:A1} and \ref{th:A1P}) take the form $\Acal=\AcalSol$. 
Accordingly, we prove such theorems by showing the inner bound $\Acal\supseteq\AcalSol$ and the outer bound $\Acal\subseteq\AcalSol$. Subsequently, we carry out the simplification $\AcalSol=\Acal_1^*$ wherever possible. Clearly, our strategy yields the classical solution for the known problems where we indeed have $\AcalSol=\Acal_1^*$. However, when
such simplification does not arise (as seen, for example, in Theorems \ref{th:A1} and \ref{th:A1P}), the classical technique fails. 

Proceeding in the same manner as in case of Theorems \ref{th:A1} and \ref{th:A1P}, we can show the inner bound $\Acal\supseteq\AcalSol$ for the known cases using the fundamental principle of source coding (given in Theorem \ref{th:funda} in the general setting and in Lemmas \ref{le:adi1} and \ref{le:adi2} in special cases). Now, from Secs.~\ref{sec:stateJ} and \ref{sec:stateP}, recall that Theorems \ref{th:A1} and \ref{th:A1P} deal with encoding under a distortion criterion and do not involve lossless coding. Also, from Secs.~\ref{sec:2Out} and \ref{sec:2pOut},  recall that, to show
the outer bound $\Acal\supseteq\AcalSol$ in 
the above cases, we required 
Shannon's lossless coding theorem and Slepian-Wolf theorem. 
Subsequently, we shall see that similar techniques will apply in case of Shannon's rate-distortion and Wyner-Ziv theorems as well, neither of which involves lossless coding. However, side information and Berger-Yeung theorems deal with problems where one source is losslessly decoded. In such cases, we shall additionally need Fano's inequality \cite{Cover}. 

Finally, consider the proof of 
$\AcalSol=\Acal_1^*$ in the four known cases at hand.
Since $$\AcalSol\supseteq \Acal^* = \bigcup_{n=1}^\infty \Acal^*_n \supseteq\Acal_1^*,$$ it is enough to show $\Acal^*_n \subseteq\Acal_1^*$ for arbitrary $n$. In fact, this has been shown in Appendix~\ref{app:specialWZ} in case of Wyner-Ziv theorem (read subscript `DC' as `WZ'). The proof makes use of convexity of $\Acal^*_1$ and reproduces crucial steps from the classical proof of Wyner-Ziv's converse theorem. Presenting this as an illustration, we omit the respective proofs of $\Acal^*_n \subseteq\Acal_1^*$ corresponding to the other known problems.
The reader with an eye for details is encouraged to construct such proofs by referring to classical proofs of converse statements of corresponding known theorems 
and
identifying 
crucial steps.

\subsection{Shannon's Lossy Coding Theorem}
\label{sec:Shan}
\label{sec:S}

{\bf {\em Formalism}:} Consider encoding of 
$X\sim p(x)$ using encoder mapping
$f:\Xcal^n \rightarrow \Zcal$ and
decoding using decoder mapping
$g: \Zcal \rightarrow \Xcal^n$ under distortion criterion $d:\Xcal^2 \rightarrow[0,\dmax]$. 
A rate-distortion pair $(R,D)$ is said to be achievable if for any $\e>0$ there exists (for $n$ sufficiently large) mapping pair $(f,g)$  such that 
\bea
\label{eq:AR}
\frac{1}{n} \log |\Zcal | &\le& R +\e \\
\label{eq:AD}
\frac{1}{n} \E d_{n}(X^{n}, \Xhat^{n}) &\le& D +\e
\eea
where $\Xhat^{n} = g(f(X^n))$.
Denote the set of achievable pairs $(R,D)$ by $\AS$. 

{\bf {\em Solution}:}
Let $\AnS^*$ be the set of $(R,D)$ pairs such that there exist alphabet $\Zcal\subseteq \Xcal^n$ and conditional distribution $q(z|x^n)$ satisfying
\bea
\label{eq:RS}
\onen
I(X^n;Z) &\le& R \\
\label{eq:DS}
\onen d_{n}(X^n,Z)&\le& D.
\eea
Moreover, denote $\AS^*=\bigcup_{n=1}^\infty \AnS^*$. Then Shannon's lossy coding theorem states $\AS =\AoneS^*$ \cite{Shannon}, whose proof according to our methodology consists of three steps showing
$\AS \supseteq \overline{\AS^*}$,
$\AS \subseteq \overline{\AS^*}$ and
$\overline{\AS^*} = \AoneS^*$, respectively. In the following, we sketch the first two steps of the proof and skip the third as indicated before. 

{\bf {\em Proof of Inner Bound}:} To show the inner bound $\AS \supseteq \overline{\AS^*}$, we generalize the argument (given immediately following the statement of Lemma \ref{le:adi1}) used in sketching the proof of $\AS \supseteq \AoneS^*$.   
For any $(R,D)\in \AS^* =\bigcup_{n=1}^\infty \AnS^*$, by definition, (\ref{eq:RS}) and (\ref{eq:DS}) hold for some $n$. Referring to Lemma \ref{le:adi1} and identifying $(Y,Z)$ with $(X^n,Z)$ and 
$R'$ with $nR$, note that (\ref{eq:RS}) is same as (\ref{eq:Rdef'}).
Consequently, by Lemma \ref{le:adi1}, for any $\e'\rightarrow 0$, there exists sequence of mapping pairs
$(f:\Xcal^{nn'} \rightarrow \Ucal,g:\Ucal \rightarrow \Zcal^{n'})$ (for some $n'\rightarrow \infty$)
such that (\ref{eq:ShRe}) and (\ref{eq:ShPr}) hold. Due to (\ref{eq:ShRe}),
$\frac{1}{nn'} \log |\Ucal|$ is close to $R$, which is same as the condition (\ref{eq:AR}), with $(\Ucal,nn')$ now playing the role of $(\Zcal, n)$. 
Further, by (\ref{eq:ShPr}), the reconstruction
$\Zhat^{n'}=g(f(X^{nn'}))$ is jointly typical with $X^{nn'}$ (the joint typicality being with respect to $(Z,X^n)$) with high probability. Hence, now dubbing 
the above reconstruction as $\Xhat^{nn'}$,
distortion $\frac{1}{nn'} \E d_{nn'}(X^{nn'},\Xhat^{nn'})$ is close to $D$ due to (\ref{eq:DS}), which is same as condition (\ref{eq:AD}) with $nn'$ in place of $n$. 
Therefore, we have $(R,D)\in \AS$, implying $\AS \supseteq \AS^*$. Further, since $\AS$ is closed, we obtain $\AS \supseteq \overline{\AS^*}$.

{\bf {\em Proof of Outer Bound}:} To show the outer bound $\AS \subseteq \overline{\AS^*}$, consider any $(R,D)\in \AS$. By definition, for any $\e>0$ there exists mapping pair $(f:\Xcal^n\rightarrow\Zcal,g:\Zcal \rightarrow \Xcal^n)$ of sufficiently large length $n$ such that (\ref{eq:AR}) and (\ref{eq:AD}) hold. Now, in view of (\ref{eq:AR}),
using interposed Shannon coding of $f(X^n)$, one can show (in the spirit of Secs.~\ref{sec:2Out} and \ref{sec:2pOut})  
\bea
\label{eq:R'S}
\onen H(g(f(X^n)))
\le
\onen
H(f(X^n))&\le& R +\e.
\eea  
Finally, for the choice $Z=g(f(X^n))$, (\ref{eq:R'S}) and (\ref{eq:AD}) become (\ref{eq:RS}) and (\ref{eq:DS}), respectively, except for $\e$ added to each right hand side.  Using an argument similar to that given in the last two paragraphs of either of
Secs.~\ref{sec:2Out} and \ref{sec:2pOut}, one can show
$(R,D)\in \overline{\AS^*}$. Hence $\AS \subseteq \overline{\AS^*}$. 

\subsection{Wyner-Ziv Theorem}
\label{sec:WZ}

{\bf {\em Formalism}:} 
Consider random variables $(X_1,X_2)\sim p(x_1,x_2)$. Encode 
$X_1$ using encoder mapping $f_{1}:\Xcal_1^n \rightarrow \Zcal_1$ 
and
decode using complete side information $X_2$ and decoder mapping
$g: \Zcal_1\times\Xcal_2^n \rightarrow \Xcal_1^n$ under distortion criterion $d:\Xcal_1^2 \rightarrow[0,\dmax]$.
A rate-distortion pair $(R_1,D)$ is said to be achievable if for any $\e>0$ there exists (for $n$ sufficiently large) a mapping pair $(f_1,g)$  such that \bea
\label{eq:ARwz}
\frac{1}{n} \log |\Zcal_1 | &\le& R_1 +\e \\
\label{eq:ADwz}
\frac{1}{n} \E d_{n}(X_1^{n}, \Xhat_1^{n}) &\le& D +\e
\eea
where $\Xhat_1^{n} = g(f_1(X_1^n),X_2^n)$.
Denote the set of achievable pairs $(R_1,D)$ by $\AWZ$. 

{\bf {\em Solution}:} 
Let $\AnWZ^*$ be the set of $(R,D)$ pairs such that there exist alphabet $\Zcal_1$, conditional distribution $q_1(z_1|x_1^n)$ and mapping $\psi:\Zcal_1\times\Xcal_2^n \rightarrow \Xcal_1^n$, satisfying
\bea
\label{eq:Rwz}
\onen
I(X_1^n;Z_1|X_2^n) &\le& R_1 \\
\label{eq:Dwz}
\onen d_{n}(X_1^n,\psi(Z_1,X_2^n))&\le& D
\eea
where $(X_1^n,X_2^n,Z_1) \sim 
p_n(x_1^n,x_2^n)
q_{1}(z_1|x_1^n)$ (i.e., $Z_1 \rightarrow X_1^n \rightarrow X_2^n$ is a Markov chain)
and $p_n(x_1^n,x_2^n) =\prod_{k=1}^n p(x_{1}(k),x_{2}(k))$. 
Further, denote $\AWZ^*=\bigcup_{n=1}^\infty \AnWZ^*$. Then Wyner-Ziv theorem states $\AWZ =\AoneWZ^*$ \cite{WZ}, whose proof according to our methodology consists of three steps showing
$\AWZ \supseteq \overline{\AWZ^*}$,
$\AWZ \subseteq \overline{\AWZ^*}$ and
$\overline{\AWZ^*} = \AoneWZ^*$, respectively. In the following, we sketch the first two steps of the proof and skip the third as indicated before. 

{\bf {\em Proof of Inner Bound}:} 
To show the inner bound $\AWZ \supseteq \overline{\AWZ^*}$, we generalize the argument (given immediately following the statement of Lemma \ref{le:adi2}) used in sketching the proof of $\AWZ \supseteq \AoneWZ^*$.
For any $(R,D)\in \AWZ^*=\bigcup_{n=1}^\infty \AnWZ^*$, by definition, 
(\ref{eq:Rwz}) and (\ref{eq:Dwz}) hold for some $n$. Referring to Lemma \ref{le:adi2} and identifying $(Y_1,Y_2,Z_1)$ with $(X_1^n,X_2^n,Z_1)$  and 
$R'_1$ with $nR_1$, note that (\ref{eq:Rwz}) is same as (\ref{eq:Rdef''}). 
Consequently, by Lemma \ref{le:adi2}, for any $\e' \rightarrow 0$, there exists sequence of mapping pairs
$(f_1:\Xcal_1^{nn'} \rightarrow \Ucal_1,g:\Ucal_1\times \Xcal_2^{nn'} \rightarrow \Zcal_1^{n'})$ (for some $n'\rightarrow \infty$)
such that
(\ref{eq:ShRe'}) and (\ref{eq:ShPr'}) hold. Due to (\ref{eq:ShRe'}),
$\frac{1}{nn'} \log |\Ucal_1|$ is close to $R_1$, which is same as the condition (\ref{eq:ARwz}) with $(\Ucal_1,nn')$ now playing the role of $(\Zcal_1, n)$. 
Further, by (\ref{eq:ShPr'}), $\Zhat_1^{n'}=g(f_1(X_1^{nn'}),X_2^{nn'})$ is jointly typical with $(X_1^{nn'},X_2^{nn'})$ with high probability. 
Hence, $X_1^{nn'}$ coupled with the reconstruction
$$\Xhat_1^{nn'} = (\psi(\Zhat_1(1),X_2^{n}(1)),\psi(\Zhat_1(2),X_2^{n}(2)),...,
\psi(\Zhat_1(n'),X_2^{n}(n')))$$
is a jointly typical sequence of $(X_1^n,\psi(Z_1,X_2^n))$ with high probability, implying, in view of (\ref{eq:Dwz}), that
the distortion $\frac{1}{nn'} \E d_{n'}(X_1^{nn'},\Xhat_1^{nn'})$ is close to $D$, which in turn is same as condition (\ref{eq:ADwz}) with $nn'$ in place $n$. 
Therefore, we have $(R,D)\in \AWZ$, implying $\AWZ \supseteq \AWZ^*$. Further, since $\AWZ$ is closed, we obtain $\AWZ \supseteq \overline{\AWZ^*}$.

{\bf {\em Proof of Outer Bound}:} To show the outer bound  $\AWZ \subseteq \overline{\AWZ^*}$, consider any $(R_1,D)\in \AWZ$. 
By definition, for any $\e>0$ there exists mapping pair $(f_1:\Xcal_1^n\rightarrow\Zcal_1,g:\Zcal_1\times \Xcal_2^n \rightarrow \Xcal_1^n)$ of sufficiently large length $n$ such that (\ref{eq:ARwz}) and (\ref{eq:ADwz}) hold. Now, in view of (\ref{eq:ARwz}),
using interposed Slepian-wolf coding of $f_1(X_1^n)$ with $X_2^n$ as the complete side information, one can show (in the spirit of Secs.~\ref{sec:2Out} and \ref{sec:2pOut})
\bea
\label{eq:R'WZ}
\onen
H(f_1(X_1^n)|X_2^n)&\le& R_1 +\e. 
\eea  
Now, for the choice $Z_1=f_1(X_1^n)$ (of course, $Z_1 \rightarrow X_1^n \rightarrow X_2^n$ is a Markov chain) and $\psi=g$, (\ref{eq:R'WZ}) and (\ref{eq:ADwz}) become (\ref{eq:Rwz}) and (\ref{eq:Dwz}), respectively, except for $\e$ added to each right hand side.  Using an argument similar to that given in the last two paragraphs of either of 
Secs.~\ref{sec:2Out} and \ref{sec:2pOut}, one can show
$(R,D)\in \overline{\AWZ^*}$. Hence $\AWZ \subseteq \overline{\AWZ^*}$.

\subsection{Side Information Theorem}
\label{sec:SI}

{\bf {\em Formalism}:} 
Consider lossless encoding using partial side information. Specifically,
encode $(X_1,X_2)\sim p(x_1,x_2)$ using encoder mapping pair
$(f_{1}:\Xcal_1^n \rightarrow \Zcal_1,f_{2}:\Xcal_2^n \rightarrow \Zcal_2)$ 
and
decode $X_1$ losslessly (in the sense of Shannon) using 
decoder mapping
$g: \Zcal_1\times\Zcal_2 \rightarrow \Xcal_1^n$. 
A rate pair $(R_1,R_2)$ is said to be achievable if for any $\e>0$, there exists (for $n$ sufficiently large) a mapping triplet $(f_1,f_2,g)$  such that \bea
\label{eq:AR1si}
\frac{1}{n} \log |\Zcal_1 | &\le& R_1 +\e \\
\label{eq:AR2si}
\frac{1}{n} \log |\Zcal_2 | &\le& R_2 +\e \\
\label{eq:APEsi}
\Pr\{X_1^{n}\ne \Xhat_1^{n}\} &\le & \e
\eea
where $\Xhat_1^{n} = g(f_1(X_1^n),f_2(X_2^n))$.
Denote the set of achievable pairs $(R_1,R_2)$ by $\ASI$. 

{\bf {\em Solution}:} 
Let $\AnSI^*$ be the set of $(R_1,R_2)$ pairs such that there exist alphabet $\Zcal_2$ and conditional distribution $q_2(z_2|x_2^n)$, satisfying
\bea
\label{eq:R1si}
\onen
H(X_1^n|Z_2) &\le& R_1 \\
\label{eq:R2si}
\onen I(X_2^n;Z_2)&\le& R_2
\eea
where $(X_1^n,X_2^n,Z_2) \sim 
p_n(x_1^n,x_2^n)
q_{2}(z_2|x_2^n)$ (i.e., $X_1^n \rightarrow X_2^n \rightarrow Z_2$ is a Markov chain)
and $p_n(x_1^n,x_2^n) =\prod_{k=1}^n p(x_{1}(k),x_{2}(k))$. 
Further, denote $\ASI^*=\bigcup_{n=1}^\infty \AnSI^*$. Then side information theorem states $\ASI =\AoneSI^*$ \cite{Wyner,AhlKor}, whose proof according to our methodology consists of three steps showing
$\ASI \supseteq \overline{\ASI^*}$,
$\ASI \subseteq \overline{\ASI^*}$ and
$\overline{\ASI^*} = \AoneSI^*$, respectively. In the following, we sketch the first two steps of the proof and skip the third as indicated before. 

{\bf {\em Proof of Inner Bound}:} For any $(R_1,R_2)\in \AnSI^*$, by definition, (\ref{eq:R1si}) and (\ref{eq:R2si}) hold for a Markov chain $X_1^n \rightarrow X_2^n \rightarrow Z_2$. Now,
consider Theorem \ref{th:funda} and recall from Sec.~\ref{sec:step3} that any rate pair 
$({R'_1},{R'_2})$, such that 
\bea
\label{eq:giri1''}
{R'_1} &\ge& I(Y_1;Z_1|Z_2)\\
\label{eq:giri2''}
{R'_2} &\ge& I(Y_2;Z_2),
\eea
also satisfies (\ref{eq:R1def})--(\ref{eq:R12def}). Further, identifying $(Y_1,Y_2,Z_1,Z_2)$ with $(X_1^n,X_2^n,X_1^n,Z_2)$ (note $I(X_1^n;Z_1|Z_2) = H(X_1^n|Z_2)$ because $Z_1=X_1^n$) and $(R'_1,R'_2)$ with  $(nR_1,nR_2)$,
 note that (\ref{eq:R1si}) and (\ref{eq:R2si})
are same as (\ref{eq:giri1''}) and (\ref{eq:giri2''}), respectively. 
Hence, (\ref{eq:R1def})--(\ref{eq:R12def}) too hold. Consequently, by Theorem \ref{th:funda}, for any $\e'\rightarrow 0$ there exists a sequence of mapping triplets $(f_1,f_2,g)$ such that (\ref{eq:rate1})--(\ref{eq:prob}) hold. Further, due to (\ref{eq:rate1}) and (\ref{eq:rate2}),
$\frac{1}{nn'} \log |\Ucal_1|$ and $\frac{1}{nn'} \log |\Ucal_2|$ are close to $R_1$ and $R_2$, respectively, which in turn are same as the respective conditions (\ref{eq:AR1si}) and (\ref{eq:AR2si}) with $(\Ucal_1,\Ucal_2,nn')$ now playing the role of $(\Zcal_1,\Zcal_2,n)$.
Recalling $Z_1=X_1^n$ and denoting $\Xhat_1^{nn'}=\Zhat_1^{n'}$, (\ref{eq:prob}) implies that $X_1^{nn'}$ differs from $\Xhat_1^{nn'}$ with low probability, which is condition (\ref{eq:APEsi}) with $nn'$ in place of $n$.
Hence $(R_1,R_2)\in \ASI$, implying $\ASI \supseteq \ASI^*$. Further, since $\AS$ is closed, we obtain $\ASI \supseteq \overline{\ASI^*}$.

{\bf {\em Proof of Outer Bound}:} The proof requires Fano's inequality (as does the classical proof of the direct statement of side information theorem), a weakened version of which states the following \cite{Cover}: Given random variables $U$ and $V$,
\be
\label{eq:Fano}
H(U|V)\le 1+\log|\Ucal|\Pr\{U\ne g(V)\}
\ee  
for any $g:\Vcal\rightarrow\Ucal$. 
Now consider any $(R_1,R_2)\in \ASI$. 
By definition, for any $\e>0$ there exists (for $n$ sufficiently large) mapping triplet
$(f_{1}:\Xcal_1^n \rightarrow \Zcal_1,f_{2}:\Xcal_2^n \rightarrow \Zcal_2,
g: \Zcal_1\times\Zcal_2 \rightarrow \Xcal_1^n)$ such that (\ref{eq:AR1si})--(\ref{eq:APEsi}) hold. Now, in view of (\ref{eq:AR1si}),
using interposed Slepian-wolf coding of $f_1(X_1^n)$ with $f_2(X_2^n)$ as the complete side information, one can show 
\bea
\label{eq:pare}
R_1 +\e &\ge&
\onen
H(f_1(X_1^n)|f_2(X_2^n))\\
\nonumber
&=& \onen H(X_1^n,f_1(X_1^n)|f_2(X_2^n)) - \onen H(X_1^n|f_1(X_1^n),f_2(X_2^n))\\
\label{eq:ek1} 
&=& \onen H(X_1^n|f_2(X_2^n)) - \onen H(X_1^n|f_1(X_1^n),f_2(X_2^n)).
\eea  
Since $\Pr\{X_1^n\ne g(f_1(X_1^n),f_2(X_2^n))\}\le \e$ by (\ref{eq:APEsi}),
we have, by Fano's inequality (\ref{eq:Fano}), 
\be
\label{eq:ek2}
H(X_1^n|f_1(X_1^n),f_2(X_2^n)) \le 1+n\e\log|\Xcal_1| \le n\e(1+\log|\Xcal_1|)
\ee
where the second inequality can be ensured by choosing $n\ge 1/\e$. Using (\ref{eq:ek2}) in (\ref{eq:ek1}), we obtain 
\be
\label{eq:ek3}
\onen H(X_1^n|f_2(X_2^n)) \le R_1 + \e (2+\log|\Xcal_1|).
\ee
Further, in view of (\ref{eq:AR2si}),
one can show, using interposed Shannon coding of $f_2(X_2^n)$, that
\be
\label{eq:ek4}
\onen H(f_2(X_2^n)) \le R_2 + \e.
\ee
Now, for the choice $Z_2=f_2(X_2^n)$ (clearly, $X_1^n \rightarrow X_2^n \rightarrow Z_2$ forms a Markov chain), (\ref{eq:ek3}) and (\ref{eq:ek4}) become (\ref{eq:R1si}) and (\ref{eq:R2si}), respectively, except for a multiple of $\e$ added to each right hand side.  Using an argument similar to that given in the last two paragraphs of either of 
Secs.~\ref{sec:2Out} and \ref{sec:2pOut}, one can show
$(R_1,R_2)\in \overline{\ASI^*}$. Hence $\ASI \subseteq \overline{\ASI^*}$. 

\subsection{Berger-Yeung Theorem}
\label{sec:BY}

{\bf {\em Formalism}:} 
Consider encoding $(X_1,X_2)\sim p(x_1,x_2)$ using encoder mapping pair
$(f_{1}:\Xcal_1^n \rightarrow \Zcal_1,f_{2}:\Xcal_2^n \rightarrow \Zcal_2)$ 
and
decoding using decoder mapping $g: \Zcal_1\times\Zcal_2 \rightarrow \Xcal_1^n\times \Xcal_2^n$. For the sake of convenience, we write $g=(g_1,g_2)$, where $g_1$ and $g_2$ has ranges that are subsets of $\Xcal_1^n$ and $\Xcal_2^n$, respectively. In particular, $g_1$ decodes
$X_1$ losslessly (in the sense of Shannon) whereas $g_2$ decodes $X_2$ under a bounded distortion criterion $d:\Xcal_2^2 \rightarrow[0,\dmax]$. 
A rate-distortion triplet $(R_1,R_2,D)$ is said to be achievable if for any $\e>0$, there exists (for $n$ sufficiently large) a mapping triplet $(f_1,f_2,g)$  such that 
\bea
\label{eq:AR1by}
\frac{1}{n} \log |\Zcal_1 | &\le& R_1 +\e \\
\label{eq:AR2by}
\frac{1}{n} \log |\Zcal_2 | &\le& R_2 +\e \\
\label{eq:APEby}
\Pr\{X_1^{n}\ne \Xhat_1^{n}\} &\le & \e\\
\label{eq:AD2by}
\frac{1}{n} \E d_{n}(X_2^{n}, \Xhat_2^{n}) &\le& D +\e
\eea
where $(\Xhat_1^{n},\Xhat_2^n) = g(f_1(X_1^n),f_2(X_2^n))$. 
Denote the set of achievable triplets $(R_1,R_2,D)$ by $\ABY$. 

{\bf {\em Solution}:} 
Let $\AnBY^*$ be the set of $(R_1,R_2,D)$ triplets such that there exist alphabet $\Zcal_2$, conditional distribution $q_2(z_2|x_2^n)$ and mapping $\psi:\Xcal_1^n\times\Zcal_2\rightarrow \Xcal_2^n$, satisfying
\bea
\label{eq:R1by}
\onen
H(X_1^n|Z_2) &\le& R_1 \\
\label{eq:R2by}
\onen I(X_2^n;Z_2|X_1^n)&\le& R_2\\
\label{eq:R12by}
H(X_1)+\onen I(X_2^n;Z_2|X_1^n)&\le& R_1+R_2\\
\label{eq:D2by}
\onen d_{n}(X_2^n,\psi(X_1^n,Z_2))&\le& D
\eea
where $(X_1^n,X_2^n,Z_2) \sim 
p_n(x_1^n,x_2^n)
q_{2}(z_2|x_2^n)$ (i.e., $X_1^n \rightarrow X_2^n \rightarrow Z_2$ is a Markov chain)
and $p_n(x_1^n,x_2^n) =\prod_{k=1}^n p(x_{1}(k),x_{2}(k))$. 
Further, denote $\ABY^*=\bigcup_{n=1}^\infty \AnBY^*$. Then Berger-Yeung theorem states $\ABY =\AoneBY^*$ \cite{BY}, whose proof according to our methodology consists of three steps showing
$\ABY \supseteq \overline{\ABY^*}$,
$\ABY \subseteq \overline{\ABY^*}$ and
$\overline{\ABY^*} = \AoneBY^*$, respectively. In the following, we sketch the first two steps of the proof and skip the third as indicated before. 

{\bf {\em Proof of Inner Bound}:} 
For any $(R_1,R_2,D)\in \AnBY^*$, by definition, (\ref{eq:R1by})--(\ref{eq:D2by}) hold for some $\psi$ and some $Z_2$ such that $X_1^n \rightarrow X_2^n \rightarrow Z_2$ is a Markov chain. Now, referring to Theorem \ref{th:funda}, let us identify $(Y_1,Y_2,Z_1,Z_2)$ with $(X_1^n,X_2^n,X_1^n,Z_2)$ and $(R'_1,R'_2)$ with  $(nR_1,nR_2)$. Noting $I(X_1^n;Z_1|Z_2) = H(X_1^n|Z_2)$ (because $Z_1=X_1^n$), (\ref{eq:R1by}) is same as (\ref{eq:R1def}). Also, (\ref{eq:R2by}) coincides with (\ref{eq:R2def}) in a straightforward manner. Next, in view of equality (\ref{eq:chain4}), we have
$$  
I(X_1^n,X_2^n;X_1^n,Z_2) = I(X_1^n;X_1^n)  + I(X_2^n;Z_2|X_1^n) = n H(X_1) + I(X_2^n;Z_2|X_1^n).
$$
Hence (\ref{eq:R12by}) coincides with (\ref{eq:R12def}).
Consequently, by Theorem \ref{th:funda}, for any $\e'\rightarrow 0$ there exists a sequence of mapping triplets $(f_1,f_2,g)$ such that (\ref{eq:rate1})--(\ref{eq:prob}) hold. Further, due to (\ref{eq:rate1}) and (\ref{eq:rate2}),
$\frac{1}{nn'} \log |\Ucal_1|$ and $\frac{1}{nn'} \log |\Ucal_2|$ are close to $R_1$ and $R_2$, respectively, which in turn are same as the respective conditions (\ref{eq:AR1by}) and (\ref{eq:AR2by}) with $(\Ucal_1,\Ucal_2, nn')$ now playing the role of $(\Zcal_1,\Zcal_2,n)$. Write $g=(g_1,g_2)$ for convenience, where ranges of $g_1$ and $g_2$ are subsets of $\Zcal_1^{n'}$ and $\Zcal_2^{n'}$, respectively. 
Recalling $Z_1=X_1^n$ and denoting $$\Xhat_1^{nn'}=\Zhat_1^{n'}=g_1(f_1(X_1^{nn'}),f_2(X_2^{nn'})),$$ (\ref{eq:prob}) implies that $X_1^{nn'}$ differs from $\Xhat_1^{nn'}$ with low probability, which is condition (\ref{eq:APEby}) with $nn'$ in place of $n$.
Again, by (\ref{eq:prob}), $\Zhat_2^{n'}=g_2(f_1(X_1^{nn'}),f_2(X_2^{nn'}))$ is jointly typical with $(\Xhat_1^{nn'},X_2^{nn'})$ with high probability. 
Hence, $X_2^{nn'}$ coupled with the reconstruction
$$\Xhat_2^{nn'} = \psi'(\Xhat_1^{nn'},\Zhat_2^{n'}) =(\psi(\Xhat_1^n(1),\Zhat_2(1)),\psi(\Xhat_1^n(2),\Zhat_2(2)),...,
\psi(\Xhat_1^n(n'),\Zhat_2(n')))$$
is a jointly typical sequence of $(X_2^n,\psi(X_1^n,Z_2))$ with high probability. Hence,
the distortion $\frac{1}{nn'} \E d_{n'}(X_2^{nn'},\Xhat_2^{nn'})$, where
$$
\Xhat_2^{nn'} = \psi'(\Xhat_1^{nn'},\Zhat_2^{n'})
=\psi'(g(f_1(X_1^{nn'}),f_2(X_2^{nn'})))=g'(f_1(X_1^{nn'}),f_2(X_2^{nn'})),
$$
 is close to $D$ due to (\ref{eq:D2by}), which in turn is same as condition (\ref{eq:AD2by}) with $(nn',g')$ in place of $(n,g)$.
Hence $(R_1,R_2,D)\in \ABY$, implying $\ABY \supseteq \ABY^*$. 
Further, since $\ABY$ is closed, we obtain $\ABY \supseteq \overline{\ABY^*}$.

{\bf {\em Proof of Outer Bound}:} 
Consider any $(R_1,R_2)\in \ABY$. 
By definition, for any $\e>0$ there exists mapping triplet
$(f_{1}:\Xcal_1^n \rightarrow \Zcal_1,f_{2}:\Xcal_2^n \rightarrow \Zcal_2,
g: \Zcal_1\times\Zcal_2 \rightarrow \Xcal_1^n\times \Xcal_2^n)$ of sufficiently large length $n$ such that (\ref{eq:AR1by})--(\ref{eq:AD2by}) hold. Now, in view of (\ref{eq:AR1by}) and (\ref{eq:AR2by}),
using interposed Slepian-wolf coding of $(f_1(X_1^n), f_2(X_2^n))$ (as seen in Sec.~\ref{sec:2Out}), one can show
\bea
\label{eq:.R1}
\onen
H(f_1(X_1^n)|f_2(X_2^n)) &\le& R_1+\e \\
\label{eq:.R2}
\onen
H(f_2(X_2^n)|f_1(X_1^n)) &\le& R_2+\e\\
\label{eq:.R12}
\onen
H(f_1(X_1^n),f_2(X_2^n)) &\le& R_1 + R_2 + 2\e.
\eea
Further, since $\Pr\{X_1^n\ne g_1(f_1(X_1^n),f_2(X_2^n))\}\le \e$ (recall $g=(g_1,g_2)$) by (\ref{eq:APEby}),
we have, by Fano's inequality (\ref{eq:Fano}), 
\be
\label{eq:ek2'}
H(X_1^n|f_1(X_1^n),f_2(X_2^n)) \le 1+n\e\log|\Xcal_1| \le n\e(1+\log|\Xcal_1|)
\ee
where the second inequality can be ensured by choosing $n\ge 1/\e$.
At this point, noting conditions (\ref{eq:.R1}) and (\ref{eq:ek2'}) to be same as conditions (\ref{eq:pare}) and (\ref{eq:ek2}), respectively, we reproduce (\ref{eq:ek3}) below: 
\be
\label{eq:ek3'}
\onen H(X_1^n|f_2(X_2^n)) \le R_1 + \e (2+\log|\Xcal_1|).
\ee
Also,
noting $f_1(X_1^n)\rightarrow X_1^n \rightarrow f_2(X_2^n)$ is a Markov chain, we have $H(f_2(X_2^n)|X_1^n)\le H(f_2(X_2^n)|f_1(X_1^n))$. Therefore, from (\ref{eq:.R2}), we have
\be
\label{eq:..R2}
\onen
H(f_2(X_2^n)|X_1^n) \le R_2+\e.
\ee
Moreover, write (\ref{eq:.R12}) as
\bea
\nonumber
R_1 +R_2 +2\e &\ge&
\onen
H(f_1(X_1^n),f_2(X_2^n))\\
\nonumber
&=& \onen H(X_1^n,f_1(X_1^n),f_2(X_2^n)) - \onen H(X_1^n|f_1(X_1^n),f_2(X_2^n))\\
\nonumber
&=& \onen H(X_1^n,f_2(X_2^n)) - \onen H(X_1^n|f_1(X_1^n),f_2(X_2^n))\\
\label{eq:ek1'} 
&=& H(X_1) + \onen H(f_2(X_2^n)|X_1^n) - \e(1+\log|\Xcal_1|)
\eea  
where (\ref{eq:ek1'}) follows using (\ref{eq:ek2'}). Rearranging (\ref{eq:ek1'}), we have
\be
\label{eq:tuki}
H(X_1) + \onen H(f_2(X_2^n)|X_1^n) \le R_1 + R_2 + \e(3 + \log|\Xcal_1|).
\ee
Further, from (\ref{eq:AD2by}), write
\bea
\nonumber
D+\e
&\ge&
\frac{1}{n} \E d_{n}(X_2^{n}, g_2(f_1(X_1^n),f_2(X_2^n))) \\
\label{eq:g2'}
&=&
\onen \E d_{n}(X_2^{n}, g_2'(\Xhat_1^n,f_1(X_1^n),f_2(X_2^n)))\\
\nonumber
&\ge& (1-\Pr\{X_1^{n}\ne \Xhat_1^{n}\}) \onen \E d_{n}(X_2^{n}, g_2'(X_1^n,f_1(X_1^n),f_2(X_2^n)))\\
\label{eq:edmax}
&\ge& \onen \E d_{n}(X_2^{n}, g_2'(X_1^n,f_1(X_1^n),f_2(X_2^n))) -\e\dmax\\
\label{eq:psix}
&=& \onen \E d_{n}(X_2^{n}, \psi(X_1^n,f_2(X_2^n))) -\e\dmax
\eea
where (\ref{eq:g2'}) holds for suitable $g_2'$, (\ref{eq:edmax}) follows because $\Pr\{X_1^{n}\ne \Xhat_1^{n}\}\le \e$ (by (\ref{eq:APEby})) and (\ref{eq:psix}) holds for suitable $\psi$. Rearranging (\ref{eq:psix}), we have
\be
\label{eq:ekn}
\onen \E d_{n}(X_2^{n}, \psi(X_1^n,f_2(X_2^n))) \le \e(1+\dmax).
\ee 
Now, for the choice $Z_2=f_2(X_2^n)$ (of course, $X_1^n \rightarrow X_2^n \rightarrow Z_2$ is a Markov chain), (\ref{eq:ek3'}), (\ref{eq:..R2}), (\ref{eq:tuki}) and (\ref{eq:ekn})
become (\ref{eq:R1by})--(\ref{eq:D2by}), respectively, except for a multiple of $\e$ added to each right hand side.  Using an argument similar to that given in the last two paragraphs of either of 
Secs.~\ref{sec:2Out} and \ref{sec:2pOut}, one can show
$(R_1,R_2,D)\in \overline{\ABY^*}$. Hence $\ABY \subseteq \overline{\ABY^*}$. 

\Section{Conclusion}
\label{sec:discuss}
In this paper, we presented a unified solution methodology that solves any two terminal source coding problem where individual encoders do not cooperate. In particular, we solved using our method the joint distortion and the partial side information problems which remained hitherto open. 
More generally, we
have shown in our analysis that
the achievable rate-distortion region in any two terminal problem admits an infinite order information-theoretic description. 
We also note that simplifications arise in certain special cases which have been known to have first order solutions. 
Summarizing,
the principal contribution of our paper lies not in providing solutions to individual source coding problems but in identifying a fundamental principle (Theorem \ref{th:funda}) of source coding arising out of the notion of typicality. In fact, this central principle captures and extends the basic typicality arguments of Shannon \cite{Shannon} and Wyner-Ziv \cite{WZ}. 
In our proofs, we also made extensive use of interposed lossless coding, which again was conceived by Wyner-Ziv. In this connection, we in turn needed Shannon's \cite{ShanLL} and Slepian-Wolf's \cite{SW} lossless coding theorems. In other words, we picked the available building blocks from existing techniques, organized them appropriately and built on them further to develop a unified framework for two terminal source coding.
In the second paper of this two part communication \cite{PartII}, we shall see that our unified framework extends to problems with any number of sources. In particular, our methodology will solve all multiterminal source coding problems where individual encoders do not cooperate. 

\newpage

\appendix
\setcounter{equation}{0}
\renewcommand{\theequation}{\Alph{section}.\arabic{equation}}

\Section{Wyner-Ziv Theorem as Corollary}
\label{app:specialWZ}
Recall that, in order to complete the derivation of Wyner-Ziv theorem as a corollary of Theorem \ref{th:A1P}, we are left to show two facts:
\begin{enumerate}

\item For a given $n$, $\AnDC^* = \{(R_1,D):(R_1,R_2,D)\in \AnDP^*, R_2 = H(X_2)\}$ is the set of $(R_1,D)$ pairs such that (\ref{eq:R1C}) and (\ref{eq:D1C}) hold for some $q_2$ and $\psi$.

\item For any $n$, $\AnDC^*\subseteq\AoneDC^*$.

\end{enumerate}
We show the above in Appendices \ref{app:WZ1} and \ref{app:WZ2}, respectively. 

\subsection{Description of \boldmath{$\Acal^*_{n\mbox{{\bf \tiny DC}}}$}}
\label{app:WZ1}

In Sec.~\ref{sec:stateP}, we have seen that  
$\AnDC^*$ is given by the set of $(R_1,D)$ satisfying (\ref{eq:R1P}) and (\ref{eq:D1P}) for some Markov chain $Z_1\rightarrow X_1^n \rightarrow X_2^n \rightarrow Z_2$. 
However, noting
$$ \{(R_1,D):(R_1,R_2,D)\in \AnDP^*\} \supseteq \{(R_1,D):(R_1,R'_2,D)\in \AnDP^*\}$$
for $R_2\ge R_2'$,
we can and shall include, without loss of generality, the additional condition
\be
\label{eq:chin}
\onen
I(X_2^n;Z_2) = H(X_2)
\ee
in the description of $\AnDC^*$. Note that (\ref{eq:chin}) implies $X_2^n=\phi(Z_2)$ with probability one for some function $\phi$, i.e., $X_1^n \rightarrow Z_2 \rightarrow X_2^n$ is a Markov chain. At the same time, recall from Sec.~\ref{sec:stateP} that we require $X_1^n \rightarrow X_2^n \rightarrow Z_2$ to form a Markov chain 
as well in the description of $\AnDC^*$. However,
the above two Markov chains can simultaneously occur in only two circumstances: 1) $X_1^n$ is independent of $(X_2^n,Z_2)$, and/or 2) $Z_2=X_2^n$ with probability one. The first possibility is ruled out assuming $X_1$ and $X_2$ are statistically dependent. Hence, accepting the second possibility and referring to (\ref{eq:R1P}) and (\ref{eq:D1P}), $\AnDC^*$ has the desired description.

\subsection{Proof of \boldmath{$\Acal_{n\mbox{{\bf \tiny DC}}}^* \subseteq \Acal_{1\mbox{{\bf \tiny DC}}}^* $}}
\label{app:WZ2}

The proof in essence reproduces part of the classical proof of Wyner-Ziv's direct theorem and assumes convexity of $\Acal_{1\mbox{{\bf \tiny DC}}}^*$.
Verify that such convexity is equivalent to convexity of the function 
$$\min_{D:(R_1,D)\in \AoneDC^*} R_1$$  
whose proof appears in \cite{WZ} and is self-contained (thus avoiding circular argument). 

For any pair $(R_1,D)\in \AnDC^*$, 
(\ref{eq:R1C}) and (\ref{eq:D1C}) hold. Next from (\ref{eq:R1C}), write:
\bea 
\nonumber
R_1 &\ge& \frac{1}{n} I(X_1^n; Z_1|X_2^n)\\
\label{eq:azx1'}
&=& \frac{1}{n}\sum_{k=1}^n I(X_{1}(k); Z_1|X_2^n,X_1^{k-1})\\
\nonumber
&=& \onen\sum_{k=1}^n H(X_{1}(k)|X_2^n,X_1^{k-1}) 
- 
\onen\sum_{k=1}^n 
H(X_{1}(k)| X_2^n,X_1^{k-1},Z_1)\\
\nonumber
&=& \onen\sum_{k=1}^n H(X_{1}(k)|X_{2}(k)) \\
\label{eq:azx2'}
&& \qquad
- 
\onen\sum_{k=1}^n 
H(X_{1}(k)| X_2^{k-1},X_{2}(k),X_{2}(k+1;n),X_1^{k-1},Z_1)\\
\label{eq:azx3'}
&\ge& \onen\sum_{k=1}^n H(X_{1}(k)|X_{2}(k)) 
- 
\onen\sum_{k=1}^n 
H(X_{1}(k)| X_2^{k-1},X_{2}(k),X_{2}(k+1;n),Z_1)\\
\label{eq:azx4'}
&=& \onen \sum_{k=1}^n  I(X_{1}(k);Z'_{1}(k)|X_{2}(k)),
\eea
where (\ref{eq:azx1'}) follows by chain rule, (\ref{eq:azx2'}) follows because $X_{1}(k)$ is independent of $X_1^{k-1}$ and by writing $X_2^{n} =
(X_2^{k-1},X_{2}(k),X_{2}(k+1;n))$, 
(\ref{eq:azx3'}) follows because conditioning reduces entropy, and (\ref{eq:azx4'}) follows by denoting $Z'_{1}(k) = (X_2^{k-1}, X_{2}(k+1;n),Z_1)$, $1\le k\le n$.
Further,
from (\ref{eq:D1C}), we can write
\be
D \ge \onen \sum_{k=1}^n \E d(X_{1}(k), \psi'_{k}(Z_1,X_2^n))
=
\frac{1}{n} \sum_{k=1}^n \E d(X_{1}(k),\psi''_{k}(Z'_{1}(k),X_{2}(k))), 
\label{eq:azx6'}
\ee
where $\psi'_{k}:\Zcal_1\times\Xcal_2^n\rightarrow \Xcal_1$
denotes the mapping that produces the $k$-th symbol produced by $\psi'$, and the mapping
$\psi''_{k}((X_2^{k-1}, X_{2}(k+1;n),Z_1),X_{2}(k)) = \psi'_{k}(Z_1,X_2^n)$
simply rearranges the arguments of $\psi'_{k}$. Comparing (\ref{eq:azx4'}) and (\ref{eq:azx6'}) with (\ref{eq:R1C}) and (\ref{eq:D1C}), respectively,
and
noting that $Z'_1(k) \rightarrow X_1(k) \rightarrow X_2(k)$ is a Markov chain for each $1\le k\le n$,  
 we have $(R_1,D) \in \AoneDC^*$ due to convexity of $\AoneDC^*$. Hence the result. \hfill $\Box$

\end{document}